\documentclass[11pt]{article}

\usepackage[utf8]{inputenc}
\usepackage{a4wide}
\usepackage{graphicx}
\usepackage{amsmath}
\usepackage{amssymb}
\usepackage{slashed}
\usepackage{listings}
\usepackage{color}
\usepackage{authblk}

\usepackage{cite}
\usepackage{fancyheadings}
\usepackage[title,titletoc]{appendix}

\numberwithin{equation}{section}

\def\preprintn{Edinburgh 2016/14}

\fancypagestyle{first}
{
  
  \fancyhead[R]{\footnotesize \preprintn}
}

\usepackage[colorlinks=true
,urlcolor=blue
,anchorcolor=blue
,citecolor=blue
,filecolor=blue
,linkcolor=blue
,menucolor=blue
,pagecolor=blue
,linktocpage=true
,pdfproducer=medialab
,pdfa=true
]{hyperref}

\newcommand{\comment}[2]{#2}

\def\A{\mathcal{A}}
\def\D{\mathcal{D}}

\def\N{\mathcal{N}}
\def\O{\mathcal{O}}
\def\s{\mathbf{s}}
\def\m{\mathbf{m}}
\def\z{\mathbf{z}}
\def\y{\mathbf{y}}
\def\N{\mathcal{N}}

\def\F{\mathbb{F}}
\def\Z{\mathbb{Z}}
\def\Q{\mathbb{Q}}

\newcommand{\mmod}[1]{\ \text{mod}\ #1}

\newcommand{\mybox}[1]{\text{\fboxsep=1.0em\fbox{$\displaystyle#1$}}}

\def\la{\langle}
\def\ra{\rangle}

\def\nn{\nonumber \\}

\title{\textbf{Scattering amplitudes over finite fields and multivariate functional reconstruction}}

\date{}

\comment{
} 

\author[]{Tiziano Peraro}
\affil[]{\emph{\small Higgs Centre for Theoretical Physics, %
School of Physics and Astronomy, %
The University of Edinburgh, %
Edinburgh EH9 3JZ, Scotland, UK }}

\begin{document}

\maketitle

\thispagestyle{first}

\begin{abstract}
  Several problems in computer algebra can be efficiently solved by
  reducing them to calculations over finite fields.  In this paper, we
  describe an algorithm for the reconstruction of multivariate
  polynomials and rational functions from their evaluation over finite
  fields.  Calculations over finite fields can in turn be efficiently
  performed using machine-size integers in statically-typed languages.
  We then discuss the application of the algorithm to several
  techniques related to the computation of scattering amplitudes, such
  as the four- and six-dimensional spinor-helicity formalism,
  tree-level recursion relations, and multi-loop integrand reduction
  via generalized unitarity.  The method has good efficiency and
  scales well with the number of variables and the complexity of the
  problem.  As an example combining these techniques, we present the
  calculation of full analytic expressions for the two-loop five-point
  on-shell integrands of the maximal cuts of the planar penta-box and
  the non-planar double-pentagon topologies in Yang-Mills theory, for
  a complete set of independent helicity configurations.
\end{abstract}

\clearpage

\tableofcontents

\section{Introduction}
\label{sec:introduction}

Scattering amplitudes in Quantum Field Theory (QFT) are an essential
ingredient for understanding the interactions among fundamental
particles that we observe in nature.  The development of techniques
and algorithms for their calculation is therefore of crucial
importance for comparing observations with theoretical predictions.
In particular, the high accuracy expected from experimental data which
is being collected at the Large Hadron Collider (LHC), as well as the
high centre-of-mass energy of the interactions produced by this
machine, require accurate predictions for processes with
high-multiplicity final states.  This has motivated, in recent years,
many studies of the structure of scattering amplitudes in QFT,
especially in gauge theories, which often led to the development of
new efficient methods for their evaluation.

Despite the remarkable recent progress in the calculation of
high-multiplicity tree-level and one-loop amplitudes, and the
numerical automation of the latter in several codes and
frameworks~\cite{Hahn:1998yk,Ossola:2007ax,Binoth:2008uq,Cullen:2011kv,Guillet:2013msa,Mastrolia:2010nb,Mastrolia:2012bu,Peraro:2014cba,Denner:2016kdg,vanHameren:2009dr,Bevilacqua:2011xh,Berger:2008sj,Hirschi:2011pa,Cullen:2011ac,Cullen:2014yla,Cascioli:2011va,Badger:2010nx,Badger:2012pg,Actis:2016mpe},
at two and higher loops these calculations are still essentially
restricted to $2 \to 2$ processes.  This is due to the significant
increase in complexity of the two-loop problem with respect to the
one-loop case.  More in detail, the most common strategy for the
calculation of loop amplitudes is to rewrite them as a linear
combination of loop integrals.  While the coefficients of this linear
combination are rational functions of kinematic invariants, the loop
integrals can instead be computed in terms of special functions.
While for low-multiplicity processes the most demanding task is
arguably the calculation of the integrals, for high-multiplicity
processes the computation of the coefficients can often have
comparable or higher complexity (notice e.g.\ that a complete planar
basis of massless five-point two-loop integrals is
known~\cite{Gehrmann:2015bfy,Papadopoulos:2015jft}, but five-point QCD
amplitudes are still unavailable for a generic helicity
configuration).  While one-loop amplitudes are often computed
numerically, two-loop amplitudes are more often computed analytically.

There are several reasons to prefer an analytic approach to a
numerical one.  Analytic expressions often yield a faster and more
stable numerical evaluation than purely numerical algorithms.
Moreover, analytic results allow to perform various kinds of studies
and manipulations, such as the analysis of the behaviour of amplitudes
in interesting kinematic limits (e.g.\ infrared and high-energy
limits).  Analytic calculations also allow to have better control over
the results and possibly infer general properties which might be
useful for the development of new analytic or numerical algorithms.

A well known bottleneck of analytic calculations in high-energy
physics is however the large size of intermediate expressions, which
can often be orders of magnitude larger than the final results.  This
is to be expected, since physical results often enjoy properties which
are not shared by each intermediate step of the calculation.
Moreover, intermediate steps are often described by a larger number of
variables (such as the loop components) which do not appear in the
final result.  The problem can be mitigated by the use of computer
algebra systems such as \textsc{Form}~\cite{Vermaseren:2000nd}, which
specializes in handling large expressions, or by using techniques such
as generalized unitarity~\cite{Britto:2004nc,Giele:2008ve}, where
intermediate steps of the calculation are gauge invariant and hence
the complexity of their expressions is reduced with respect to a
diagrammatic approach.

In this paper we assess the possibility of side-stepping the issue of
large intermediate expressions, by reconstructing analytic results
from their numerical evaluation, where each intermediate step is
trivially a number or a set of numbers (this will be better defined in
the next paragraph).  A polynomial or a rational function can be
reconstructed, with very high probability, from its numerical
evaluation at several values of its arguments.  In particular we focus
on the functional reconstruction of multivariate polynomials and
rational functions with applications to calculations in high-energy
physics.

The first question to address is which kind of numerical evaluation is
suited for a functional reconstruction.  An obvious choice would be a
floating-point calculation, but these are affected by numerical
inaccuracies which would add an additional layer of complexity to the
functional reconstruction algorithms.  Exact calculations might
instead be performed over the field of rational numbers.  However,
numerical calculations with rational numbers are affected by a similar
problem to the one of analytic calculations.  Indeed, while a large
intermediate expression would be translated into a rational number, in
general the number of digits of the numerator and the denominator of
this number would be very high.  This requires extensive use of
computationally expensive arbitrary-precision arithmetic, which can
significantly slow down the calculation.  A common and successful
approach in computer algebra is the use of finite fields, which have a
finite number of elements and can be represented by machine-size
integers, offering the possibility of performing fast but exact
calculations in statically-typed languages such as \textsc{C} and
\textsc{C++}.  Their main drawback is that, switching from the
rational field to a finite field, some information is lost and must be
recovered by repeating a functional reconstruction over several finite
fields.  This strategy is however much more efficient than a
calculation over the rational field.

The usage of finite fields in computer algebra in actually quite
common.  Many computer algebra systems use finite fields under the
hood for solving problems such as polynomial factorization and
Greatest Common Divisor (GCD).  The application of finite fields in
high-energy physics has been introduced in
ref.~\cite{vonManteuffel:2014ixa}, in the context of
Integration-By-Parts (IBPs).  However, to the best of our knowledge,
an application of finite-field techniques to a realistic problem in
high-energy physics involving the reconstruction of multivariate
rational functions is not present in the literature and is presented
here for the first time.  In particular, the main missing ingredient
which we illustrate in this paper is the application of a functional
reconstruction algorithm capable of handling relatively complex
results which depend on many variables, such as those appearing in
typical high-multiplicity multi-loop calculations.

The paper is roughly divided in two parts.  In the first part we
describe dense\footnote{ A \emph{dense} reconstruction algorithm for
  polynomials and rational functions, unlike a \emph{sparse}
  reconstruction algorithm, seeks to be efficient in the general case
  where the result has many non-vanishing terms, rather than in the
  special case where it only has a small number of non-vanishing terms
  compared with its total degree.}  functional reconstruction
algorithms for univariate and multivariate polynomials and rational
functions.  These algorithms reconstruct polynomials and rational
functions from their repeated numerical evaluation over finite fields
(although in principle they can actually be used over any field) and
they are independent of the specific algorithm used for their
evaluation.  In particular, for univariate polynomials we use the well
known Newton's polynomial representation.  For univariate rational
functions, we use Thiele's interpolation formula.\footnote{ The
  formula is named after the mathematician Thorvald Nicolai Thiele
  (1838--1910).}  For multivariate polynomials, we use a recursive
version of Newton's formula.  For multivariate rational functions we
were not able to find a dense reconstruction algorithm suited for our
needs in the literature.  However we found that the technique proposed
in ref.~\cite{CUYT20111445} for sparse rational functions can be
easily adapted to the dense case, by combining it with the other
techniques we mentioned for univariate rational functions and
multivariate polynomials.  The resulting algorithm is capable of
efficiently reconstructing functions with many non-vanishing terms and
depending on several variables, as we will show in the examples.
Using these methods, the analytic calculation of any polynomial or
rational function can be turned into the problem of providing an
efficient numerical evaluation of the same function over finite
fields.

The second part of the paper concerns the application of the mentioned
reconstruction algorithms to techniques relevant for the calculation
of scattering amplitudes.  It should be stressed that any algorithm
which can be implemented via a sequence a rational elementary
operations (addition, subtraction, multiplication and division) is
suited for the usage of these reconstruction techniques, which
therefore have a very broad spectrum of applications.  In particular,
widely used methods such as tensor reduction and IBPs obviously fall
into this category.  In this paper we however focus on multi-loop
integrand reduction via generalized
unitarity~\cite{Britto:2004nc,Giele:2008ve,Ossola:2006us,Mastrolia:2011pr,Badger:2012dp,Zhang:2012ce,Mastrolia:2012an,Badger:2012dv,Kleiss:2012yv,Feng:2012bm,Mastrolia:2012wf,Mastrolia:2013kca,Badger:2013gxa,Mastrolia:2016dhn,Kosower:2011ty,Larsen:2012sx,CaronHuot:2012ab,Johansson:2012zv,Johansson:2013sda,Johansson:2015ava,Ita:2015tya,Larsen:2015ped},
since the algorithm is suited for high-multiplicity processes and, as
stated, writing scattering amplitudes as linear combinations of
integrals (which may be further processed by IBPs at a later stage) is
currently one of the main bottlenecks of high-multiplicity multi-loop
calculations.  These techniques have indeed been used in recent five-
and six-point calculations of two-loop amplitudes in
non-supersymmetric Yang-Mills
theory~\cite{Badger:2013gxa,Badger:2015lda,Badger:2016ozq}.  In order
to provide the building blocks needed by generalized unitarity, we
also discuss in some detail a finite-field implementation of the
spinor-helicity formalism in four~\cite{Mangano:1987xk,Berends:1987me}
and six dimensions~\cite{Cheung:2009dc,Bern:2010qa,Davies:2011vt}, as
well as the calculation of tree-level amplitudes over finite fields
via Berends-Giele recursion~\cite{Berends:1987me}.  In particular, the
six-dimensional spinor-helicity formalism is used to provide a
higher-dimensional embedding of loop momenta and spinors, which will
thus have an explicit finite-field numerical representation.

As explicit examples combining all these techniques, we present the
calculation of full analytic expressions for the two-loop five-point
on-shell integrands of the maximal cuts of the planar penta-box and
the non-planar double-pentagon topologies in Yang-Mills theory, for a
complete set of independent helicity configurations.  In particular,
in the all-plus case we find agreement with the results available in
the literature, while for the other helicity configurations the result
is new and we make it publicly available for comparisons with future
calculations.

All the algorithms discussed in the paper have been implemented in a
\textsc{C++} library which can produce the mentioned analytic results
from finite-field evaluations using 64-bit integers, without relying
on any external computer algebra system.

The paper is structured as follows.  In sect.~\ref{sec:notation} we
set the notation and review some notions about finite fields which we
will use in the rest of the paper.  While all these notions are well
known, they are rarely used in high-energy physics and hence they are
reviewed in some detail for the convenience of the reader.  In
sect.~\ref{sec:funct-reconstr} we describe the functional
reconstruction algorithms mentioned above.  In
sect.~\ref{sec:spinor-helicity-tree} we discuss the implementation of
spinor-helicity and tree-level techniques over finite fields.  These
are mostly meant as a stepping stone for the discussion of integrand
reduction and generalized unitarity, which is illustrated in
sect.~\ref{sec:multi-loop-integrand}, where we also provide the
two-loop five-point examples mentioned above.  In
sect.~\ref{sec:implementation} we give some details about our
proof-of-concept \textsc{C++} implementation of the algorithms
illustrated in this paper, which might be useful for other
implementations.  In sect.~\ref{sec:conclusions-outlook} we finally
draw our conclusions and briefly discuss further possible applications
of these techniques.  In Appendix~\ref{sec:basic-finite-field} we
recall some well known theorems and algorithms involving modular
arithmetic and finite fields, highlighting the role they play in the
reconstruction algorithms illustrated in this paper.  More details
about our usage of the six-dimensional spinor-helicity formalism over
finite fields are given in Appendix~\ref{sec:six-dimens-momenta}.  In
Appendix~\ref{sec:two-loop-unitarity} we discuss an efficient method
for generating two-loop unitarity cuts from Berends-Giele currents,
which is a generalization of the one-loop algorithm used by the public
code \textsc{NJet}~\cite{Badger:2012pg}, and whose two-loop extension
is not present elsewhere in the literature.

\section{Basic concepts, definitions and notation}
\label{sec:notation}

In this section we set the notation and review some well known
concepts about finite fields which will be used later in the paper.

\subsection{Finite fields}
\label{sec:finite-fields}

Finite fields are fields containing a finite number of elements.  For
the purposes of this paper, we will only consider fields of integers
modulo $p$, denoted by $\Z_p$, where $p$ is a prime number.

In particular, we identify $\Z_n$, with $n$ a positive integer (not
necessarily prime), with the set of non-negative integers\footnote{
  One could alternatively define $\Z_n$ as a set of equivalence
  classes in $\Z$, but we find the definition given in this section
  more pragmatic and useful for the purposes of this paper.} smaller
than $n$,
\begin{equation}
  \Z_n = \{0,\ldots,n-1\}.
\end{equation}
Elementary arithmetic operations in $\Z_n$, such as addition,
subtraction and multiplication, are defined using modular arithmetic,
namely by performing the corresponding operation in $\Z$ and taking
the remainder of the integer division of the result modulo $n$.  Given
an element $a\in\Z_n$ with $a\neq 0$, if $a$ and $n$ are co-prime we
can define the inverse $a^{-1}$ of $a$ in $\Z_n$ with respect to
multiplication, i.e.\ an element $b\in Z_n$ such that
\begin{equation}
  a^{-1} \mmod{n} \equiv b \qquad \Leftrightarrow \qquad (a\, b) \mmod{n} = 1.
\end{equation}
One can indeed show that such a number $b$ exists (and is unique in
$\Z_n$) if and only if $a$ and $n$ are co-prime.  If $n=p$ is a prime
number, the existence of an inverse is therefore guaranteed for every
non-vanishing element of $\Z_p$.  This implies that $\Z_p$ is a field
and any rational operation on its elements is well defined.  The
multiplicative inverse can be computed using the extended Euclidean
algorithm, as explained in Appendix~\ref{sec:mult-inverse}.

The existence of a multiplicative inverse implies that we can define a
map between rational numbers and elements of a finite field $\Z_p$.
In particular, given a rational number $q=a/b\in\Q$, we define
\begin{equation}
  q\mmod{p} \equiv \big( a\times (b^{-1} \mmod{p}) \big) \mmod{p}.
\end{equation}
This map is obviously not invertible (since $\Q$ is infinite and
$\Z_p$ is finite), however it turns out one can reconstruct $q$, with
very high probability, from its images in several finite fields
$\Z_{p_i}$ where $\{p_i\}$ is a set of prime numbers, as explained in
sect.~\ref{sec:rati-reconstr} and
Appendix~\ref{sec:basic-finite-field}.  This will enable us to
reconstruct rational functions with rational coefficients from their
values over finite fields.  It is worth observing that we can
similarly map $q$ in $\Z_n$, with $n$ not prime, as long as $n$ and
the denominator $b$ are co-prime.

\subsection{Polynomials and rational functions}
\label{sec:polyn-rati-funct}

In this paper we use a multi-index notation.  Given a sequence of
$n$ variables $\z=(z_1,\ldots,z_{n})$, and the $n$-dimensional
multi-index $\alpha=(\alpha_1,\ldots,\alpha_n)$ with integers
$\alpha_i\geq 0$, a monomial $\z^\alpha$ is defined as
\begin{equation}
  \z^\alpha \equiv \prod_{i=1}^n z_i^{\alpha_i}.
\end{equation}
The total degree of a monomial is denoted by $|a|$,
\begin{equation}
  |a| = \sum_i \alpha_i.
\end{equation}
With $\F$ a generic field, we use the following (standard)
definitions:
\begin{itemize}
\item $\F[\z]$ is the ring of polynomials in the variables $\z$ with
  coefficients in $\F$.  Any polynomial function $f\in \F[\z]$ can be
  uniquely identified by a set of multi-indexes $\{\alpha\}$ and
  coefficients $c_\alpha\in\F$ as
  \begin{equation} \label{eq:defpoly}
    f(\z) = \sum_\alpha c_\alpha\, \z^\alpha.
  \end{equation}
\item $\F(\z)$ is the field of rational functions in the variables
  $\z$ with coefficients in $\F$.  Functions $f\in \F(\z)$ can be
  expressed as a ratio of two polynomials $p,q\in\F[\z]$ as
  \begin{equation} \label{eq:defrational}
    f(\z) = \frac{p(\z)}{q(\z)} = \frac{\sum_\alpha n_\alpha\, \z^\alpha}{\sum_\beta d_\beta\, \z^\beta},
  \end{equation}
  where $n_\alpha,d_\beta\in\F$, while $\{\alpha\}$ and $\{\beta\}$
  are sets of multi-indexes.  Unlike the polynomial representation
  given in Eq.~\eqref{eq:defpoly}, the representation of a rational
  function is not unique, even if we assume $p$ and $q$ to have no
  common polynomial factors.  However, after GCD simplification, the
  only possible ambiguity is an overall constant normalization of the
  numerator and the denominator.  In order to have a unique
  representation, useful e.g.\ when comparing functions obtained in
  different finite fields, we use the convention of defining the
  coefficient of the lowest degree term of the denominator (with
  respect to the chosen monomial order) to be equal to one.
\end{itemize}

The functional reconstruction methods described in this paper are
based on multiple evaluations of the function $f$ to be reconstructed,
which in turn correspond to assigning to each variable a value in the
field $\F$.  For the univariate case we denote these values by
$y_i\in\F$ while multivariate we denote them by $\y_i\in \F^n$, where
$i$ is a label distinguishing different evaluation points.

In this paper the field $\F$ will either be the rational field $\Q$ or
a finite field $\Z_p$.  More in detail, our goal is the calculation of
polynomial functions in $\Q[\z]$ and rational functions in $\Q(\z)$.
We will do so by performing a functional interpolation of the same
functions in $\Z_p[\z]$ or $\Z_p(\z)$ respectively, for several primes
$p$ if needed, and then use these to reconstruct the results over the
rational field.

\subsection{Rational reconstruction from finite fields}
\label{sec:rati-reconstr}

In the next sections we will describe an algorithm for efficiently
reconstructing polynomials and rational functions over finite fields.
As apparent from their representation in Eq.~\eqref{eq:defpoly}
and~\eqref{eq:defrational}, these functions can be identified by a
sequence of monomials and their respective coefficients.  The final
step of the reconstruction algorithm therefore consists in promoting
these coefficients from elements of a finite field $\Z_p$ into a
proper rational number.  As we have seen in
sect.~\eqref{sec:finite-fields}, one can map a rational number $q$
into an element of the set $\Z_n$ of integers modulo $n$, as long as
$n$ and the (reduced) denominator of $q$ are co-prime.  Although this
map is not invertible, it turns out one can use a variation of the
extended Euclidean algorithm~\cite{Wang:1981:PAU:800206.806398} (see
also Appendix~\ref{sec:mult-inverse}) in order to make a guess for $q$
from its image $q\mmod{n}\in \Z_n$.  This method is known as
\emph{rational reconstruction}.  The guess will, in general, be
correct when the numerator and the denominator of $q$ are much smaller
than $n$ (heuristically one finds that the threshold is around
$\sqrt{n}$).  However, because of our requirement of working with
machine-size integers (see Sect.~\ref{sec:implementation} for more
details), we cannot always choose the prime $p$ defining the field
$\Z_p$ to be significantly larger than the numerator and the
denominator of any rational number which can be expected to appear in
the results.

The solution to the problem comes from the Chinese remainder theorem,
which allows to uniquely reconstruct an element in $\Z_n$, with
$n=n_1\cdots n_k$ and the $n_i$ pairwise co-prime, from its images in
$\Z_{n_i}$ for $i=1,\ldots k$, as explained in
Appendix~\ref{sec:chinese-remainder}.  This implies that, given a set
of prime numbers $p_i$, we can perform the functional reconstruction
over $\Z_{p_i}$ and then combine the results to find the image of the
rational coefficients in $\Z_{p_1\cdots p_k}$.  By performing the
rational reconstruction over $\Z_{p_1\cdots p_k}$, the result will
thus be correct when the product of the selected primes is large
enough.\footnote{ A minor subtlety arises when the denominator of a
  rational number is a multiple of one of the primes $p_i$.  We
  observe that this is very unlikely to happen if the $p_i$ are of
  $\O(10^{6})$ as in our implementation.  Besides, in this case the
  functional reconstruction would fail and thus one can simply discard
  the prime and proceed with a different one.}  This allows us to use
machine-size integers for a fast functional reconstruction in
$\Z_{p_i}$, while the use of multi-precision arithmetic
(computationally much more expensive) is restricted to this rational
reconstruction step in $\Z_{p_1\cdots p_k}$, which takes a very small
fraction of computing time compared to the one spent for the
functional reconstruction over prime fields.

More in detail, given a sequence of primes $p_1,\ldots,p_k$, we adopt
the following algorithm for the full reconstruction of a rational
function in $f\in\Q(\z)$ (a completely analogous one can obviously be
used for a polynomial):
\begin{enumerate}
\item[1.] Reconstruct the function $f$ in $\Z_{p_1}(\z)$ and store the
  result.
\item[2.] Use the rational reconstruction algorithm to promote the
  stored result to a (new) guess $g\in\Q(\z)$.
\item[3.] Consider a new prime $p_{i+1}$, where $i$ is the number of
  primes which have already been used so far.  Evaluate the guess $g$
  over the new field $\Z_{p_{i+1}}$ for several values of $\z$.  If
  the result for $g(\z)$ agrees with the evaluation of the function
  $f(\z)$, accept the guess $g$ as the correct answer, i.e.\ assert
  $g(\z)=f(\z)$, and successfully terminate the algorithm.  Otherwise
  proceed to the next point.
\item[4.] Reconstruct the function in $\Z_{p_{i+1}}(\z)$ and combine
  it with the stored result in $\Z_{p_1\cdots p_i}(\z)$ using the
  Chinese remainder theorem in order to obtain the correct result in
  $\Z_{p_1\cdots p_{i+1}}(\z)$.  The latter thus replaces the
  previously stored result.  Repeat from point $2$.
\end{enumerate}
The algorithm terminates when the comparison in point 3 is successful.
For the examples presented in this paper, we typically only need to
perform the functional reconstruction over one or two prime fields.

We observe that the techniques reviewed in this section (which, as
stated, are well known) allow reconstructing a multivariate rational
function with very high probability.  In practice, exceptional cases
are very artificial and irrelevant for realistic applications.  It
should also be noted that the final result can be extensively checked
against the evaluation of the function $f$ on even more values of the
prime $p$ and the variables $\z$.

In the rest of this paper, we will discuss a functional reconstruction
algorithm which is suited for complex theoretical calculations in
high-energy physics, and its application to techniques related to the
computation of tree-level and multi-loop scattering amplitudes in QFT.

\section{Functional reconstruction}
\label{sec:funct-reconstr}

In this section we describe a dense functional reconstruction
algorithm for polynomials and rational functions whose performance
scales well with the complexity of the result.  For the sake of
generality we make no further assumption about the functions to be
reconstructed.  Unless explicitly stated otherwise, the techniques
illustrated in this section can be applied to functions over any
field, although in practice we are mostly interested to their
application over finite fields.

\subsection{The black-box interpolation problem}
\label{sec:black-box-interp}

Given a function $f$ of $n$ variables $\z=(z_1,\ldots,z_n)$ over a
field $\F$, the so-called \emph{black-box interpolation problem} is
the problem of reconstructing the function $f$ from the results of its
evaluation for several values of $\z$.  In other words, one can think
about the function as a numerical procedure of the form
\begin{equation} \label{eq:blackbox}
  \z \longrightarrow \mybox{f} \longrightarrow f(\z),
\end{equation}
while the functional reconstruction method has no knowledge about the
algorithm used for the calculation of $f$.

In this paper we are interested in reconstructing polynomials and
rational functions over the rational field.  Our setup is a modified
version of the black-box interpolation problem, where the function is
evaluated modulo a prime number $p$, and can schematically be
represented as
\begin{equation} \label{eq:blackboxmodp}
  (\z,p) \longrightarrow \mybox{f} \longrightarrow f(\z) \mmod{p}.
\end{equation}
We recall that the rational reconstruction technique reviewed in
sect.~\ref{sec:rati-reconstr} reduces the problem of a functional
reconstruction over $\Q$ to the problem of a functional reconstruction
over prime fields $\Z_p$ for generic $p$.  This makes the setups in
Eq.~\eqref{eq:blackbox} and Eq.~\eqref{eq:blackboxmodp} effectively
equivalent for our purposes.  Hence, in the remainder of this paper,
we will focus on functional reconstruction techniques over finite
fields $\Z_p$, where $p$ is an arbitrary prime.  As stated, these are
actually valid in any field $\F$ and the use of finite fields is meant
to provide a fast but exact numerical evaluation of the black-box
function $f$ to be reconstructed.  Therefore, in this section we will
simply denote the generic field by $\F$.

The advantage of turning an analytic calculation into a black-box
interpolation is that it reduces the problem of computing a function
$f$ into the one of providing a fast numerical evaluation for it.
Since the reconstruction is independent of the algorithm used for the
evaluation of $f$, it has a very broad spectrum of applications.
Numerical calculations can in turn avoid issues such as large
intermediate expressions, which affect many computations in
high-energy physics.  With this approach, the number of evaluations
needed for the reconstruction of a function scales linearly with the
number of terms of the result itself and is independent of the
complexity of intermediate expressions which may appear using fully
analytic techniques.

We remind the reader that, because we are dealing with polynomials and
rational functions, which can be represented as in
Eq.~\eqref{eq:defpoly} and Eq.~\eqref{eq:defrational} respectively,
the goal of a functional reconstruction algorithm is to identify the
monomials appearing in their definition and the corresponding
coefficients as elements of the field $\F$.

Functional reconstruction algorithms roughly fall into two categories:
\emph{dense} and \emph{sparse} algorithms.  As suggested by their
name, sparse reconstruction algorithms attempt to be more efficient
(with respect to number of function evaluations needed) in the case
where the number of non-vanishing terms is small compared to the one
expected from their total degree.  In this paper we focus, as
mentioned, on dense reconstruction algorithms, seeking good efficiency
in the most general case where many non-vanishing terms are present,
rather than being optimal in the special cases where the function to
be reconstructed is relatively simple.

In order to better motivate the discussion which follows, it is useful
to consider first a straightforward system-solving strategy for the
functional reconstruction, and point out why it is not suitable for
our purposes.  Given a set of values $\y_i\in \F^n$, using
Eq.~\eqref{eq:defpoly} and Eq.~\eqref{eq:defrational}, one can build
systems of equations of the form
\begin{equation}
  \sum_\alpha c_\alpha\, \y_i^\alpha - f(\y_i) = 0
\end{equation}
and
\begin{equation}
    \sum_\alpha n_\alpha\, \y_i^\alpha  - \sum_\beta d_\beta\, \y_i^\beta f(\y_i) = 0,
\end{equation}
to be solved for the coefficients $\{c_\alpha\}$ and
$\{n_\alpha,d_\beta\}$ for polynomials and rational functions
respectively, using basic techniques such as Gauss elimination.
Notice, however, that one cannot know a priori which monomials will
appear in the result.  Moreover, in sect.~\ref{sec:polyn-rati-funct}
we defined the coefficients of our canonical representation of
rational functions such that the coefficient of the lowest degree term
in the denominator is equal to one, in order to solve the ambiguity in
their representation.  Unfortunately we have no knowledge about the
mentioned term before having performed the reconstruction.  This is
however a minor issue which will be solved as discussed in
sect.~\ref{sec:mult-rati-funct} using techniques proposed in
ref.~\cite{CUYT20111445}.  The same techniques will allow us to assess
the total degree of the numerator and the denominator of a
multivariate rational function, or the one of a polynomial, using a
relatively small number of evaluations.  Hence, a viable solution to
the functional reconstruction problem consists in listing the full set
of $N$ monomial terms compatible with the total degree of the function
involved, sampling the function with (at least) $N$ values for the set
of variables $\z$, and solving the resulting $N\times N$ system of
equations for the coefficients.  While this method is straightforward
and efficient for simple functions depending on only one or two
variables, it has however a bad scaling behaviour when increasing the
number of variables or the total degree of the result.  This can be
understood simply by recalling that solving an $N\times N$ dense
system of linear equations is an $\O(N^3)$ operation, and the
multivariate problems in which we are interested in can have several thousands
(or even hundreds of thousands) of potentially non-vanishing terms
(notice e.g.\ that the most general polynomial in $n$ variables and
total degree $R$ has $N={R+n \choose R}$ terms).

The reconstruction algorithms we are going to describe have instead a
much better scaling with the complexity of the result, and the time
spent for the functional reconstruction itself is typically much
smaller than the time required for evaluating the function to be
reconstructed.  In the following, we start by describing well
established algorithms for the univariate case and later use them as
ingredients for the multivariate one.

\subsection{Univariate polynomials}
\label{sec:univ-polyn}

For univariate polynomial functions $f=f(z)$ we adopt a well known
reconstruction method based on Newton's polynomial representation.
Given a sequence of distinct values $y_0,\ldots,y_R \in \F$, a
univariate polynomial $f\in\F[z]$ of total degree $R$ can be written
as~\cite{abramowitz1964handbook}
\begin{align} \label{eq:unewtonrep}
  f(z) ={}& \sum_{r=0}^R a_r \prod_{i=0}^{r-1} (z-y_i) \nn
         = {}& a_0 + (z-y_0)\bigg(a_1 + (z-y_1)\Big( a_2 + (z-y_2) \big(\cdots+ (z-y_{r-1})\, a_r\big) \Big) \bigg).
\end{align}
The coefficients $a_r$ can be computed recursively by evaluating the
function $f$ at the values $y_i$ as
\begin{align} \label{eq:newtonsystem}
  a_0 ={}& f(y_0) \nn
  a_1 = {} & \frac{f(y_1)-a_0}{y_1-y_0} \nn
  a_2 = {} & \Big(\big(f(y_2)-a_0\big) \frac{1}{y_2-y_0}-a_1\Big)\frac{1}{y_2-y_1}\nn
  \cdots = {} & \cdots \nn
  a_r = {} & \Bigg(\Big(\big(f(y_r)-a_0\big) \frac{1}{y_r-y_0}-a_1\Big) \frac{1}{y_r-y_1}-\cdots - a_{r-1} \Bigg) \frac{1}{y_r-y_{r-1}}.
\end{align}
An important feature of the method is that each coefficient $a_r$ only
depends on the evaluation of $f$ at the points $y_i$ with $i\leq r$.
This implies that new evaluation points cannot change the value of the
previously computed coefficients.  This is ideal for the case where
the total rank $R$ of the polynomial $f$ is not known a priori, and it
is our main reason for preferring this method over alternatives.  In
practice, we apply the algorithm recursively until we find a set of
consecutive coefficients $a_r$ which evaluate to zero.  This is the
termination criterion of the algorithm (notice that, even when the
canonical form of the polynomials has several vanishing entries, in
general the entries of its Newton representation will be
non-vanishing, hence the described termination criterion is
robust and an incorrect termination is extremely unlikely).

The sequence ${y_i}$ is generated dynamically, taking into account
that the algorithm evaluating of the function $f$ might fail at a
particular point.  More in detail, we choose $y_0$ as an arbitrary
element of the field $\F$ and then we recursively define the following
ones as $y_i=y_{i-1}+1$, as long as the evaluation of $f$ is
successful.  If the evaluation of $f$ fails, we try replacing the
current point $y_i$ with $y_i+1$, until we find a value for which the
evaluation of $f$ is possible.  If too many consecutive evaluations
fail, we terminate the algorithm declaring the reconstruction
unsuccessful.

Even though Newton's representation is more practical for functional
reconstruction purposes, after a succesful reconstruction it is
convenient to convert it back into the canonical representation given
by Eq.~\eqref{eq:defpoly}, which in the univariate case can be written
as
\begin{align} \label{eq:defupoly}
  f(z) = \sum_{r=0}^R c_r \, z^r.
\end{align}
It is worth observing that the conversion from the representation in
Eq.~\eqref{eq:unewtonrep} to the one in Eq.~\eqref{eq:defupoly} only
requires the following two operations
\begin{itemize}
\item addition of univariate polynomials,
\item multiplication of a univariate polynomial by a linear univariate
  polynomial.
\end{itemize}
Both operations are simple enough to be efficiently implemented over
finite fields $\Z_p$ in statically-typed languages, without resorting
to external computer algebra systems.  More details on our
implementation are given in sect.~\ref{sec:implementation}.

\subsection{Univariate rational functions}
\label{sec:univ-rati-funct}
For univariate rational functions $f\in \F(z)$ it is worth
distinguishing two cases.  The first case is applicable when the
degree of the numerator and the denominator ($R$ and $R'$
respectively) are known and the constant term of the denominator is
known to be non-vanishing.  As we shall see, this is useful for the
multivariate reconstruction discussed in
sect.~\ref{sec:mult-rati-funct}.  In this particular case, the
function admits a representation of the form
\begin{equation} \label{eq:defuratfun}
  f(z) = \frac{\sum_{r=0}^R n_r\, z^r}{\sum_{r'=0}^{R'} d_{r'}\, z^{r'}},
\end{equation}
with $d_0=1$.  We find the system-solving method explained at the end
of section~\ref{sec:black-box-interp} is well suited for this
univariate case, since the rank of the numerator and the denominator
are unlikely to be high enough to make Gauss elimination impractically
expensive, at least for the kind of problems we are interested in this
paper (as we stated, the same is not true for the multivariate case,
as the number of variables or the degree of the numerator and the
denominator increase).  Having set $d_0=1$, the system of equations
reads
\begin{equation}
  \sum_{r=0}^R n_r\, y_i^r - \sum_{r'=1}^{R'} d_{r'}\, y_i^{r'} f(y_i) = f(y_i),
\end{equation}
and can be solved for the unknown coefficients $\{n_r,d_{r'}\}$ by
evaluating the function $f$ at least $R+R'+1$ times.  In practice we
include a few more evaluation points $y_i$, making the system slightly
over-constrained as a cross check.

We now address the more general case where the degree of the numerator
and the denominator are not known.  We find convenient to use a method
based on Thiele's interpolation formula~\cite{abramowitz1964handbook},
which expresses a rational function $f\in\F(z)$ as a continued
fraction
\begin{align} \label{eq:thielerep}
  f(z) = {} & a_0 + \dfrac{z-y_0}{a_1 + \dfrac{z-y_1}{a_2 + \dfrac{z-y_3}{\cdots + \dfrac{z-y_{r-1}}{a_N}}}} \nn
       = {} & a_0 + (z-y_0)\left(a_1 + (z-y_1)\left( a_2 + (z-y_2) \left(\cdots+ \frac{z-y_{N-1}}{a_N}\right)^{-1} \right)^{-1} \right)^{-1},
\end{align}
where $y_0,\ldots,y_{N}$ is a sequence of distinct elements of $\F$.
Thiele's interpolation formula can be regarded as the analog of
Newton's formula for rational functions.  The second line of
Eq.~\eqref{eq:thielerep}, by comparison with the second line of
Eq.~\eqref{eq:unewtonrep}, makes the analogy manifest.

The coefficients $a_i$ can be recursively computed by evaluating the
function $f$ at the values $y_i$,
\begin{align} \label{eq:thielesolve}
  a_0 ={}& f(y_0) \nn
  a_1 = {} & \frac{y_1-y_0}{f(y_1)-a_0} \nn
  a_2 = {} & \left(\left(f(y_2)-a_0\right)^{-1}(y_2-y_0)-a_1\right)^{-1} (y_2-y_1)\nn
  \cdots = {} & \cdots \nn
  a_j = {} & \left(\left(\left(f(y_j)-a_0\right)^{-1}(y_j-y_0)-a_1\right)^{-1} (y_j-y_1)-\cdots -a_{j-1}\right)^{-1} (y_j-y_{j-1}).
\end{align}
Similarly to Newton's interpolation formula, each value $a_j$ only
depends on the evaluations at $z=y_i$ with $i\leq j$, which makes the
algorithm convenient in the case where the total number of terms is
not known, since new evaluation points do not change previously
computed terms.  Our strategy for generating the sequence of points
$y_i$ is also the same as the one used in the polynomial case.  It is
worth pointing out that Thiele's interpolation formula, being a
continued fraction, contains spurious singularities which might make
the application of equations \eqref{eq:thielesolve} impossible.
Similarly to the case where the evaluation of $f$ fails, if such a
spurious singularity is encountered at a point $y_i$, we simply
discard the value and replace it with $y_i+1$.  The termination
criterion is the agreement between a new evaluation $f(y_i)$ with the
evaluation of the rational function defined by the coefficients
$a_0,\ldots a_{i-1}$ already computed.  We make the algorithm more
robust by requiring agreement between the reconstructed function and
several new evaluations of $f$.

After a successful interpolation, the result is converted into the
canonical form given by Eq.~\eqref{eq:defuratfun}, except that the
condition $d_0=1$ is replaced by $d_{\min r'}=1$, where $z^{\min r'}$
is the lowest degree monomial with a non-vanishing coefficient in the
denominator.  We observe that Thiele's formula with $N+1$ terms
represents a rational function with degree $R$ and $R'$ for the
numerator the denominator respectively, where $R=R'=N/2$ if $N$ is
even, and $R=R'+1=(N+1)/2$ if $N$ is odd.  In other words, either the
degree of the numerator and the denominator of the reconstructed
function are equal, or they differ by one unity at most.  This implies
that the highest degree coefficients $n_r$ or $d_{r'}$, obtained by
converting the result into its canonical form, might be vanishing, in
which case they are discarded.  For this reason, if the degrees $R$
and $R'$ are already known, the system-solving strategy typically
requires fewer evaluations of the function, since this way one can
avoid reconstructing zeros.  Thiele's interpolation is however
preferred when $R$ and $R'$ are not known.

The conversion into a canonical form can be implemented by performing
on the polynomial numerator and the denominator of the function the
same kind of operations we listed for converting Newton's polynomial
representation into the canonical form (and inversion, which is simply
implemented by swapping numerator and denominator).  Hence, as we
stated for the previous case, this does not require the usage of
computer algebra systems and can be easily implemented in
statically-typed languages using functions over finite fields $\Z_p$.

We stress that the result we obtain can be shown to be minimal with
respect to the degrees of the numerator and the denominator, and hence
no GCD simplification is needed after the reconstruction is converted
into a canonical form (and possible high-degree zero terms are
discarded).

\subsection{Multivariate polynomials}
\label{sec:mult-polyn}

Given a sequence of variables $\z=(z_1,\ldots,z_n)$ and a multivariate
polynomial function $f\in \F[\z]$, the interpolation of $f$ can be
performed by recursive application of the univariate Newton's
reconstruction method described above.  In other words, we consider a
generic multivariate polynomial $f\in \F[z_1,\ldots,z_n]$ as a
univariate polynomial in the first variable $z_1$, whose coefficients
are polynomials in the remaining variables $(z_2,\ldots, z_n)$.
Newton's interpolation formula~\eqref{eq:mnewtonrep} can thus be
generalized to the multivariate case, by upgrading the coefficients
$a_i$ from elements of the field $\F$ to elements of the polynomial
ring $\F[z_2,\ldots,z_n]$,
\begin{equation} \label{eq:mnewtonrep}
  f(z_1,\ldots,z_n) ={} \sum_{r=0}^R a_r(z_2,\ldots,z_n) \prod_{i=0}^{r-1} (z_1-y_i),
\end{equation}
where, as before, the $y_i$ are distinct elements of $\F$.  The
solutions for the coefficients $a_r$ in Eq.~\eqref{eq:newtonsystem}
also apply to the multivariate case, with the following substitutions
\begin{equation}
  f(y_j) \longrightarrow f(y_j,z_2,\ldots,z_n), \qquad a_j \longrightarrow a_j(z_2,\ldots,z_n).
\end{equation}
In this case, the right hand side of each
equation~\eqref{eq:newtonsystem} for $a_r(z_2,\ldots,z_n)$ thus
depends on the previously computed polynomial coefficients
$a_{j}(z_2,\ldots,z_n)$ with $j<r$, and on the function $f$ where the
first variable $z_1$ has been set to the value $z_1=y_r$.  Therefore
the reconstruction of $a_r(z_2,\ldots,z_n)$ is reduced to the
black-box interpolation of a polynomial function in $n-1$ variables.
In particular, the evaluation of $a_r(z_2,\ldots,z_n)$ can be obtained
by evaluating $f(y_r,z_2,\ldots,z_n)$ and then combining it with the
previously computed polynomial coefficients $a_j$ by applying the
formulas in Eq.~\eqref{eq:newtonsystem}.  The recursion ends with the
univariate case, where we apply the univariate polynomial
reconstruction algorithm we already discussed.

In order to convert a multivariate polynomial from a (recursive)
Newton representation into the canonical form in
Eq.~\eqref{eq:defpoly}, one needs the following basic operations
\begin{itemize}
\item addition of multivariate polynomials,
\item multiplication of a multivariate polynomial by a linear
  univariate polynomial.
\end{itemize}
Although these operations are slightly more involved than the
corresponding ones for the univariate case, they still allow a rather
straightforward and efficient implementation in statically-typed
languages, especially in the case of polynomials over finite fields
$\Z_p$.  We will discuss the details of our implementation and the
polynomial representation we used in sect.~\ref{sec:implementation}.

\subsection{Multivariate rational functions}
\label{sec:mult-rati-funct}

The reconstruction of multivariate rational functions is a
considerably more complex problem that those addressed so far.  We
were not able to find a dense multivariate reconstruction algorithm
suited for our purposes in the literature.  However, we observe that
the techniques proposed in ref.~\cite{CUYT20111445} for sparse
rational functions, can also be applied, with some minor
modifications, to dense rational functions as well.  In this section
we describe our version of the techniques proposed
in~\cite{CUYT20111445}, and in particular their combination with the
dense functional reconstruction methods discussed above.  The
efficiency of the resulting algorithm turns out to meet our
performance goals, as we shall see in the next sections.

We consider the variables $\z=(z_1,\ldots,z_n)$ and a rational
function $f\in\F(\z)$, which admits a canonical representation of the
form of Eq.~\eqref{eq:defrational}.  A first issue to be addressed is
the ambiguity in the overall normalization of the numerator and
denominator of the function.  As we stated, we get rid of this
ambiguity by requiring the canonical representation of the function to
have the coefficient of the lowest degree term in the denominator
equal to one.  However, identifying such a term (or indeed any term in
the function) before the functional reconstruction is in general not
possible.  As observed in ref.~\cite{CUYT20111445}, there is however a
simple case where this issue can be easily solved, i.e.\ when the
lowest-degree non-vanishing term of the denominator is the constant
term, in which case we can simply set the canonical normalization by
imposing
\begin{equation} \label{eq:md0eq1}
  d_{(0,\ldots,0)}=1.
\end{equation}
It is also noted in ref.~\cite{CUYT20111445} that, even in the case
where $d_{(0,\ldots,0)}=0$, one can always identify a shift
$\s=(s_1,\ldots,s_n)\in\F^n$ in the arguments of $f$ such that the
shifted function
\begin{equation} \label{eq:fshift}
  f_\s(\z) = f(\z+\s) = f(z_1+s_1,\ldots,z_n+s_n)
\end{equation}
satisfies $d_{(0,\ldots,0)}\neq 0$ (i.e.\ is non-singular in
$z=(0,\ldots,0)$) and can thus be normalized as in
Eq.~\eqref{eq:md0eq1}.  Hence, in these cases, the method we are going
to describe can be applied to the function $f_\s$ instead.  A minor
subtlety is the fact that, after such a shift, the new function $f_\s$
might become considerably more complex than the original function $f$,
and thus harder to reconstruct.  This issue is actually much more
relevant for the sparse reconstruction case discussed in
ref.~\cite{CUYT20111445}, but it can also affect the dense
reconstruction case, especially if the variables $\z$ have been
carefully chosen for being suited to describe the function at hand.
We will address this subtlety later and for the time being we turn to
the description of the functional reconstruction of a rational
function $f\in\F(\z)$ whose constant term in the denominator is
non-vanishing.

The method illustrated in ref.~\cite{CUYT20111445} consists in
introducing an auxiliary variable $t$ and defining a new function
$h\in \F(t,\z)$ as
\begin{equation}
  h(t, \z) = f(t\, \z) = f(t\, z_1, \ldots, t\, z_n).
\end{equation}
Using the canonical representation of $f$ given by
Eq.~\eqref{eq:defrational} and denoting the total degree of the
numerator and the denominator of $f$ by $R$ and $R'$ respectively, we
get
\begin{equation} \label{eq:ratrech}
  h(t,\z) = \frac{\displaystyle\sum_{r=0}^R p_r(\z)\, t^r}{1+\displaystyle\sum_{r'=1}^{R'} q_{r'}(\z)\, t^{r'}},
\end{equation}
where
\begin{align}
  p_r(\z) \equiv{} \sum_{|\alpha|=r} n_\alpha\, \z^\alpha, \qquad
  q_{r'}(\z) \equiv{} \sum_{|\beta|=r'} d_\beta\, \z^\beta.
\end{align}
In other words, the function $h$ can be regarded as a univariate
rational function in the variable $t$, whose coefficients are
multivariate homogeneous polynomials in the variables $\z$.  We
obviously have
\begin{equation} \label{eq:ratrecuh}
  f(\z) = h(1,\z).
\end{equation}
Notice that the constant term of the denominator of $f$ coincides with
the constant term of the denominator of $h$ seen as a univariate
rational function in $t$, hence by normalizing the latter to 1, the
same normalization is automatically applied to the former.

Our implementation of the method consists in reconstructing the
homogeneous polynomials $p_r$ and $q_{r'}$ for $0\leq r\leq R$ and
$1\leq r' \leq R'$, using the multivariate Newton reconstruction
method illustrated above.  The evaluations of these polynomials at a
generic point $\z=\y_i\in\F^n$ are in turn obtained by reconstructing
\begin{equation}
  h_{\y_i}(t) \equiv h(t,\y_i)
\end{equation}
as a univariate rational function in $t$ and identifying $p_r(\y_i)$
and $q_{r'}(\y_i)$ with its coefficients.  Notice that this univariate
reconstruction is actually equivalent to the evaluation of all the
polynomials $p_r$ and $q_{r'}$ at the same time.  In order not to
waste evaluations, for each sampled value $\y_i \in \F^n$ we cache the
corresponding reconstructed univariate function $h_{y_i}\in\F(t)$ so
that its coefficients can be re-used for the reconstruction of several
polynomials.  Moreover, since $p_r(\z)$ and $q_{r'}(\z)$ are
homogeneous polynomials of degree $r$ and $r'$ respectively, during
their reconstruction we can drop their dependence on the first
variable $z_1$ by setting $z_1=1$ and consider the polynomials
$p_r(1,z_2,\ldots,z_n)$ and $q_{r'}(1,z_2,\ldots,z_n)$ instead.  The
dependence on $z_1$ is thus restored by homogenizing the result.  This
simplification makes up for having introduced the auxiliary variable
$t$.

We now briefly discuss the univariate reconstruction algorithm we use
for the functions $h_{\y_i}\in\F(t)$ at each point $\y_i \in \F^n$.
Since we typically cannot know a priori the degrees $R$ and $R'$ of
the numerator and the denominator respectively, we first perform a few
univariate reconstructions using Thiele's interpolation method, in
order to obtain this information.  As a byproduct, this is also used
to check whether the constant term of the denominator vanishes, in
which case, as mentioned, we proceed by specifying a different shift
$\z\to \z+\s$ in the arguments of $f$ (more details about this point
are given below).  Once the degree of the numerator and the
denominator are known, and a shift $\s$ such that
$d_{(0,\ldots,0)}\neq 0$ is found, we switch to the system-solving
algorithm described at the beginning of
sect.~\ref{sec:univ-rati-funct}, since it typically requires fewer
evaluations of the function (as explained, Thiele's interpolation
becomes optimal with respect to the number of evaluations needed only
when the degree of the numerator and the denominator are equal or the
former is one unity higher than the latter).

We finally briefly describe a convenient way to implement the
aforementioned variable shift to avoid the case where
$d_{(0,\ldots,0)}\neq 0$ (this point is more extensively discussed in
ref.~\cite{CUYT20111445}, to which we refer the reader for further
details).  For simplicity we focus on the polynomials $p_r$ in the
numerator of the function, although it should be clear that completely
analogous statements can be made for the polynomials $q_r$ in the
denominator.  The key observation is that a shift in the variables
$\z\to \z+\s$ in one of the homogeneous polynomials $p_r$ can only
affect the homogeneous polynomials $p_{r'}$ with $r'<r$.  In
particular the highest degree polynomial $p_R$ is the same for the
function $f$ and the shifted function $f_\s$ defined in
Eq.~\eqref{eq:fshift}.  We then start with the reconstruction of the
highest degree polynomial $p_R$, which is the same for $f$ and $f_\s$.
We then compute the effects of the shift on lower rank terms, which
for a generic $p_r$ are given by
\begin{equation}
  p_r(\z+\s)-p_r(\z).
\end{equation}
During the reconstruction of lower degree terms, each function
evaluation of the polynomials defining $f_\s$ are thus corrected by
the effects of this shift induced by higher rank terms.  With this
method we can reconstruct the function $f$ directly (up to an overall
common normalization of the numerator and the denominator, which can
be fixed at the end using our definition of their canonical form),
which as stated can be expected to be simpler than $f_\s$.

As a final remark, we point out that, on top of the basic polynomial
operations needed for the multivariate polynomial reconstruction, in
order to implement this algorithm one also needs to compute the
effects of a shift $\z\to\z+\s$ in the variables.  This is relatively
straightforward if done for one variable at a time, by using
\begin{equation}
  (z+s)^r = \sum_{j=0}^r {r \choose j} s^{r-j}\, z^j.
\end{equation}
Another operation which is needed is the homogenization required to
restore the dependence of the polynomials $p_r$ and $q_r$ on their
first variable, but depending on the polynomial representation used
this can be trivial, since it amounts to adjusting the first entry of
the multi-indexes $\alpha=(\alpha_1,\ldots,\alpha_n)$ representing the
exponents of the monomials.  As already mentioned, more details about
our implementation are given in sect.~\ref{sec:implementation}.

\bigskip

This completes the description of the functional reconstruction
algorithms used in this paper.  In
sections~\ref{sec:spinor-helicity-tree}
and~\ref{sec:multi-loop-integrand} we show how to evaluate functions
relevant for calculations high-energy physics over finite fields
$\Z_p$, making thus possible to use these techniques for the
reconstruction of their analytic form as rational functions over the
field $\Q$.

\subsection{Examples}

Before moving to applications in high-energy physics, we briefly
discuss two simple examples of functional reconstruction, which might
be useful in order to clarify the concepts introduced in this section,
especially for the reconstruction of multivariate rational functions,
which is considerably more involved than the other cases.  We perform
the reconstruction over the field $\Q$, since it is easier to follow,
but any step can equivalently be performed over finite fields $\Z_p$,
yielding the same result modulo $p$.

\subsubsection*{Function non-singular at $\z=(0,\ldots,0)$}
We consider the following rational function of two variables
\begin{equation}
  f(\z) = f(z_1,z_2) = \frac{3 + 2 z_1 + 4 z_2 + 7 z_1^2 + 5 z_1 z_2 + 6 z_2^2}{1 + 7 z_1 + 8 z_2 + 10 z_1^2 + z_1 z_2 + 9 z_2^2}.
\end{equation}
Because the function is well defined at $\z=(0,0)$ there is no need to
shift variables in this case.  Our goal is thus to reconstruct the
analytic formula of this function from a black-box providing a
numerical evaluation of it.  The first step consists in defining the
function $h$ in Eq.~\eqref{eq:ratrech}, which in this case is given by
\begin{equation} \label{eq:ratrechex1}
  h(t,\z) = f(t\, z_1,t\, z_2) = \frac{p_0 + p_1(z_1,z_2)\, t^2 + p_2(z_1,z_2)\, t^2}{1 + q_1(z_1,z_2)\, t^2 + q_2(z_1,z_2)\, t^2},
\end{equation}
with
\begin{alignat}{2} \label{eq:ratfunpqex1}
  p_0(z_1,z_2)={}& 3, \qquad \qquad & q_0(z_1,z_2) \equiv {} & 1, \nn
  p_1(z_1,z_2)={}& 2 z_1 + 4 z_2, \qquad \qquad  & q_1(z_1,z_2) = {}&  z_1 + z_2, \nn
  p_2(z_1,z_2)={}& 7 z_1^2 + 5 z_1 z_2 + 6 z_2^2, \qquad \qquad  & q_2(z_1,z_2) ={} & 10 z_1^2 + z_1 z_2 + 9 z_2^2.
\end{alignat}
We thus reconstruct the polynomials $p_r$ and $q_{r'}$ by
interpolating the function $h_{\y_i}(t)\equiv h(t,\y_i)$ for several
points $\y_i$.  As stated, since the polynomials $p_r$ and $q_{r'}$
are homogeneous and have degree $r$ and $r'$ respectively, we can set
$z_1=1$ and homogenize the result at the end.

We assume no knowledge about the total degree of $f$.  Hence
we first perform a reconstruction of $h_{\y_1}(t)$ using Thiele's
interpolation formula for a specific value of $\y_1$, say
$\y_1\equiv(1,1)$, which yields
\begin{equation}
  h_{(1,1)}(t) =\frac{3 + 6 t + 18 t^2}{1 + 2 t + 20 t^2}.
\end{equation}
This gives us information about the total degree of the numerator and
the denominator, which can then be confirmed by repeating the
reconstruction at more values of $\z=\y_i$.  Moreover, by comparison
with Eq.~\eqref{eq:ratrechex1}, this also yields the value of the
polynomials $p_r$ and $q_{r'}$ at $(z_1,z_2)=(1,1)$, namely
\begin{alignat}{2}
  p_0(1,1)={}& 3, \qquad \qquad & q_0(1,1) = {} & 1, \nn
  p_1(1,1)={}& 6, \qquad \qquad  & q_1(1,1) = {}&  2, \nn
  p_2(1,1)={}& 18, \qquad \qquad  & q_2(1,1) ={} & 20,
\end{alignat}
which can be used for their polynomial fit.  We perform two more
reconstructions, say at $\y_2=(1,2)$ and $\y_3(1,3)$, which yield
\begin{align}
  h_{(1,2)}(t) ={}& \frac{3+10 t + 41 t^2}{1+ 3 t+ 48 t^2} \nn
  h_{(1,3)}(t) ={}& \frac{3+14 t+76 t^2}{1+4 t+ 94 t^2}.
\end{align}
These, together with the previous point $\z=\y_1$, are sufficient in
order to reconstruct, using e.g.\ Newton's formula, the univariate
polynomials
\begin{alignat}{2}
  p_1(1,z_2)={}& 2 + 4 z_2, \qquad \qquad  & q_1(1,z_2) = {}& 1 + z_2, \nn
  p_2(1,z_2)={}& 7 + 5 z_2 + 6 z_2^2, \qquad \qquad  & q_2(1,z_2) ={} & 10 + z_2 + 9 z_2^2.
\end{alignat}
After homogenization, these reproduce the ones in
Eq.~\eqref{eq:ratfunpqex1} and thus the function $f(\z)=h(1,\z)$.

\subsubsection*{Function singular at $\z=(0,\ldots,0)$}
We now briefly discuss the case of a function which is singular at
$\z=(0,0)$.  We consider
\begin{equation} \label{eq:ratrecex2}
  f(\z) = f(z_1,z_2) = \frac{3 + 2 z_1 + 4 z_2 + 7 z_1^2 + 5 z_1 z_2 + 6 z_2^2}{z_1 + z_2 + 10 z_1^2 + z_1 z_2 + 9 z_2^2},
\end{equation}
which differs from the previous example only for the missing constant
term in the denominator.  Using Thiele's interpolation formula for
$h_{(1,1)}(t)$, we now get
\begin{align}
  h_{(1,1)}(t) ={} \frac{3/2+3 t+9 t^2}{t+10 t^2},
\end{align}
which tells us that the constant term of the denominator is vanishing
and thus we need to specify an argument shift.  By trial and error,
repeating the univariate functional reconstruction for several shifts,
we find e.g.\ that $\s=(2,0)$ is a good shift, i.e.\
\begin{equation}
  f_{\s}(\z) = f(z_1+2, z_2)
\end{equation}
is non-singular at $(z_1,z_2)\equiv 0$.  The function $h_\s$ is now
defined as
\begin{equation}
  h_{\s}(t,\z) = f(t\, z_1+2, t\, z_2)
\end{equation}
and can similarly be written as in Eq.~\eqref{eq:ratrechex1}.  Now its
shifted polynomial coefficients are
\begin{alignat}{2}
  p_{\s,0}(z_1,z_2)={}& \frac{35}{43}, \qquad & q_{\s,0}(z_1,z_2) \equiv {} & 1, \nn
  p_{\s,1}(z_1,z_2)={}& \frac{30}{43}  z_1 + \frac{14}{43} z_2, \qquad  & q_{\s,1}(z_1,z_2) = {}& \frac{41}{43} z_1 + \frac{3}{43} z_2, \nn
  p_{\s,2}(z_1,z_2)={}& \frac{7}{43} z_1^2 + \frac{5}{43} z_1 z_2 + \frac{6}{43} z_2^2, \qquad  & q_{\s,2}(z_1,z_2) ={} & \frac{10}{43} z_1^2 + \frac{1}{43} z_1 z_2 + \frac{9}{43} z_2^2.
\end{alignat}
These coefficients can be reconstructed similarly to the previous
case, but as we explained we use a different strategy where the
effects of the shift are recursively subtracted from the evaluations
of each polynomial coefficient.  This allows us to reconstruct the
polynomials coefficients $p_r$ and $q_{r'}$ of the original function
$f$, up to an overall normalization of the numerator and the
denominator, rather than the coefficients $p_{\s,r}$ and $q_{\s,r'}$
of the shifted function $f_\s$.  For simplicity we focus on the
reconstruction of the polynomials $p_r$ in the numerator.  We observed
that the highest degree polynomial $p_2$ is independent of the chosen
shift, and thus we can simply reconstruct it from several
interpolations of $h_{\s,\y_i}(t)$ as in the non-singular case.  We
then compute the effects of the variable shift on the lower degree
terms, which are given by
\begin{equation}
  p_2(z_1+2,z_2)-p_2(z_1,z_2) = \Big(\frac{28}{43} z_1 + \frac{10}{43} z_2\Big) + \Big(\frac{28}{43}\Big).
\end{equation}
The term in the first and second parenthesis on the r.h.s.\ are the
effects of the variable shift coming from $p_2$ on the coefficients
$p_1$ and $p_0$ respectively.  When reconstructing $p_1$, we thus
evaluate $p_{\s,1}$ (for which we can re-use the interpolations of
$h_{\s}$ we already performed for $p_2$) and correct its values using
\begin{equation}
  p_1(z_1,z_2) = p_{\s,1}(z_1,z_2)-\Big(\frac{28}{43} z_1 + \frac{10}{43} z_2\Big) = \frac{2}{43} z_1+\frac{4}{43} z_2.
\end{equation}
Hence the polynomial $p_{\s,1}$ is never actually reconstructed, since
we correct its evaluation as in the previous equation and reconstruct
$p_{1}$ instead.  We thus proceed with the reconstruction of $p_0$,
for which we need the effects of the shift on $p_1$
\begin{equation}
  p_1(z_1+2,z_2)-p_1(z_1,z_2) = \Big(\frac{4}{43}\Big),
\end{equation}
which combined with the ones on $p_2$ give
\begin{equation}
  p_0 = p_{\s,0}-\frac{4}{43}-\frac{28}{43} = \frac{3}{45}.
\end{equation}
Once again, the evaluation of $p_{\s,0}$ is obtained by the
interpolations of $h_\s$ we already performed for $p_2$.  A similar
strategy can obviously be applied to the denominator.  While $q_2$,
following the same arguments, is not affected by the variable shift
(i.e.\ $q_2=q_{\s,2})$, the other terms corrected by the shift are
\begin{equation}
  q_2(z_1,z_2) = \frac{z_1}{43}+\frac{z_2}{43}, \qquad q_0 = 0.
\end{equation}
Notice that, after correcting for the shift, the lowest degree term of
the denominator, namely $z_2/43$, does not have coefficient equal to
unity.  In order to bring the result into our canonical form, all the
coefficients are thus multiplied by its inverse (in this case $43$),
obtaining the result in Eq.~\eqref{eq:ratrecex2}.

\section{Spinor-helicity and tree-level techniques}
\label{sec:spinor-helicity-tree}

In this section we describe how spinor-helicity techniques and
tree-level calculations can be implemented over finite fields $\Z_p$,
and thus used in combination with the reconstruction algorithms
illustrated above.  In particular, one has to ensure that both the
inputs and each intermediate step of a calculation performed via these
technique can be implemented through a sequence of rational
operations.  These techniques also serve as building blocks for
multi-loop calculations via integrand-reduction and generalized
unitarity, whose implementation on finite fields will be described in
sect.~\ref{sec:multi-loop-integrand}.

More in detail, in sect.~\ref{sec:four-dimens-momenta} we describe how
to use the well known four-dimensional spinor-helicity formalism in
finite-field calculations.  As we shall see, for loop calculations in
the context of generalized unitarity, it can also be convenient to
embed loop momenta in a $\D$-dimensional space with integer $\D>4$.
For this purpose, we will also use the six-dimensional spinor-helicity
formalism~\cite{Cheung:2009dc,Bern:2010qa,Davies:2011vt}, as described
in Appendix~\ref{sec:six-dimens-momenta}.  In
sect.~\ref{sec:berends-giele-recurs} we recall how to combine these
techniques for tree-level calculations by means of Berends-Giele
recursion.

\subsection{Four-dimensional momenta and spinors}
\label{sec:four-dimens-momenta}

The four-dimensional spinor-helicity formalism offers a very
convenient way of expressing helicity amplitudes in gauge theories.
This formalism is widely used and here we only review a few details
which are relevant for our implementation.

Given a massless momentum $p$, its corresponding two-component spinors
$|p\ra$ and $|p]$ can be defined as the independent solutions of the
Dirac equation,
\begin{equation}
  p^\mu\, \sigma_\mu |p\ra = p^\mu\, \sigma_\mu |p] = 0,
\end{equation}
where $\sigma_\mu = (\mathbf{1}_{2\times 2},\sigma_i)$ with $\sigma_i$
being the Pauli matrices, and they satisfy
\begin{equation}
 |p\ra\, [p| =  p_\mu\, \sigma^\mu.
\end{equation}
In particular, if $p$ is the momentum of an outgoing particle of a
process, $|p\ra$ and $|p]$ are identified as the negative- and
positive-helicity solution of the Dirac equation respectively.

We consider an arbitrary $n$-point process, with external momenta
$p_1,\ldots,p_n$, all taken as outgoing for simplicity.  In the
following we assume all external particles are massless, but our
statements can be easily generalized to the massive case by performing
a massless decomposition of the massive external momenta into a sum of
massless vectors.  For each external particle $p_i$ we have the
associated spinors $|p_i\ra$ and $|p_i]$, also denoted by $|i\ra$ and
$|i]$ for simplicity.  Scattering amplitudes are rational functions of
the spinor components, as apparent from well known relations valid for
external momenta
\begin{equation} \label{eq:sp4momenta}
  p^\mu = \frac{1}{2}\, \la p |\, \sigma^\mu\,  | p ],
\end{equation}
and polarization vectors
\begin{equation} \label{eq:sp4polvec}
  \epsilon_+^\mu(p,\eta) = \frac{\la \eta | \sigma^\mu | p]}{\sqrt{2}\, \la \eta\, p \ra}, \qquad \epsilon_-^\mu(p,\eta) = \frac{\la p | \sigma^\mu | \eta]}{\sqrt{2}\, [ p\, \eta ]},
\end{equation}
where $\eta$ is a reference vector such that the scalar product $(p\cdot \eta) \neq 0$.

The total number of these spinor components is however much larger
than the number of independent variables needed to describe an
amplitude.  We therefore find much more convenient to adopt the
strategy described in
references~\cite{Hodges:2009hk,Badger:2013gxa,Badger:2016uuq},
which consists in describing scattering amplitudes with a minimal
number of variables, which in turn provide a rational parametrization
of the components of the spinors.

More in detail, we recall that, under a redefinition of the spinor
phases given by the little group tranformation
\begin{equation} \label{eq:littlegroup}
  |i\ra \to t_i\, |i\ra , \qquad |i] \to \frac{1}{t_i}\, |i],
\end{equation}
an $n$-point amplitude $\A(1,\ldots,n)$ transforms as
\begin{equation}
  \A(1,\ldots,n) \to \left(\prod_{i=1}^n t_i^{-2\, h_i} \right) \A(1,\ldots,n),
\end{equation}
where $h_i$ is the helicity of the $i$-th particle (e.g.\ $\pm 1/2$
for fermions and $\pm 1$ for gluons).  It is thus convenient to
extract from the amplitude an overall factor
$\A^{\textrm{(phase)}} (1,\ldots,n)$ which, under
Eq.~\eqref{eq:littlegroup}, has the same transformation properties as
the amplitude $\A$.  While there is obviously no unique choice for
$\A^{\textrm{(phase)}}$, it is clear that it can be chosen based on
the external helicities only and does not require any other knowledge
about the amplitude $\A$ itself (see e.g.\ ref.~\cite{Badger:2016uuq}
for a choice valid for any number of external legs, or
sect.~\ref{sec:2lpb} for a five-point example).

The ratio $\A/\A^{\textrm{(phase)}}$ is obviously phase-free, i.e.\
invariant under the transformation in Eq.~\eqref{eq:littlegroup}.  Any
phase-free function of the (four-dimensional) kinematic components
which is invariant under the Poincaré group can depend on $3\, n-10$
independent invariants.  Depending on the problem at hand, these may
be chosen as a subset of the Mandelstam invariants
$s_{ij}=(p_i+p_j)^2$, however these are not suited for our purposes
since they don't provide a rational parametrization of the spinor
components.  It is however not hard to build, for an arbitrary
process, a set of $3\, n-10$ invariants which, up to a Poincaré and
little group transformation, yield a rational parametrization of the
spinor components.  A particularly convenient one, based on earlier
work by Hodges on momentum twistors~\cite{Hodges:2009hk}, is given in
references~\cite{Badger:2013gxa,Badger:2016uuq} (we will also give some
examples later, when discussing explicit applications).  Hence, in the
following we denote by $x_1,\ldots,x_{3 n -10}$ a complete set of
invariants which provide a rational parametrization of the spinor
components (up to a Poincaré and little group transformation).  We
refer to these as \emph{momentum-twistor variables}.  Twistor
variables can be interpreted as a rational parametrization of the
phase space and, despite having been introduced in the context of
conformal theories, they can be used in any relativistic quantum field
theory and indeed they played an important role in multi-leg
higher-loop calculations in non-supersymmetric gauge theories
presented recent
years~\cite{Badger:2013gxa,Badger:2015lda,Badger:2016ozq}.  Since they
can give a complete description of the ratio
$\A/\A^{\textrm{(phase)}}$, a generic scattering amplitude can be
rewritten in terms of them as
\begin{equation}
  \A(1,\ldots,n) = \A^{\textrm{(phase)}}(1,\ldots,n)\, \tilde \A (x_1,\ldots,x_{3 n - 10}).
\end{equation}
Because $\tilde \A$ is a rational function of the momentum-twistor variables,
it is suited to being evaluated over finite fields $\Z_p$, and
therefore to the application of the functional reconstruction
algorithms described earlier in this paper.

When dealing with helicity amplitudes, our starting point is therefore
the definition of the spinor components in terms of momentum-twistor variables.
In practice we observe that we can always choose our variables such
that all but one, say $x_1$, are dimensionless.  For the purpose of
the functional reconstruction, it is thus convenient to set $x_1=1$
and recover its functional dependence at the end via dimensional
analysis.  Hence our chosen set of independent variables is
\begin{equation}
  \z = \left(x_2,\ldots,x_{3 n - 10}\right).
\end{equation}
If internal or external massive particles are involved, we should also
add their masses to the list of the independent variables.  From the
components of the spinors, we then build the external momenta $p^\mu$.
Explicit analytic formulas for these in terms of the corresponding
spinors can be easily worked out once and for all using
Eq.~\eqref{eq:sp4momenta}.  In particular we represent any (massless
or massive) momentum $p$ by its light-cone components
\begin{equation} \label{eq:4dmomenta}
  p \equiv (p^0+p^3,\ p^0-p^3,\ p^1+i\, p^2,\ p^1-i\, p^2),
\end{equation}
where, with respect to their usual definition, we dropped a factor
$1/\sqrt{2}$, which being irrational cannot be translated into
$\Z_p$.\footnote{ In principle square roots and factors $i$ may also
  be included by considering more general finite fields, but since
  these are always overall factors which cancel out in the final
  result for $\tilde \A$, we do not need to introduce this additional
  complication into the algorithm.}  With this representation, the
light-cone components of $p$ are also rational functions of the
momentum-twistor variables (factors $i$ cancel out against those in
the Pauli matrices).  Similarly, we compute polarization vectors
according to Eq.~\eqref{eq:sp4polvec}, except that we divide them by
an extra $\sqrt{2}$ (more details are given in the next paragraph).
The resulting expression, in light-cone components, is also a rational
function of the momentum-twistor variables.

A minor issue arises from overall factors $i$ and $\sqrt{2}$ present
in the definition of polarization vectors and in the color-ordered
Feynman rules of gauge theories.  These factors are however completely
spurious, since $\tilde \A(x_1,\ldots,x_{3 n - 10})$ is always a
rational function of the momentum-twistor variables, provided that a
factor $i$ is also extracted in $\A^{\text{(phase)}}$.  Therefore, if
$\epsilon^\mu$ is an external polarization vector, $V_n$ is a generic
$n$-point vertex, and $P$ is a propagator, with colour matrices
normalized as $\textrm{tr}[T^a,T^b]=\delta^{a b}$, we get rid of these
unwanted factors by making the following substitutions
\begin{equation}
  \epsilon^\mu \to \frac{1}{\sqrt{2}}\, \epsilon^\mu, \qquad V_n \to (-i\,  2\sqrt{2})^{4-n}\, V_n, \qquad P \to i\, P.
\end{equation}
Notice that the rescaling of $V_n$ is consistent when combining
different kinds of vertexes (e.g.\ $V_3\times V_3$ scales in the same
way as
$V_4$).  The correct overall factor, which is a rational number, is
restored at the end with a power counting on the number of external
gluons, internal vertexes and propagators.

The ingredients outlined so far are sufficient for expressing a
tree-level amplitude, as well as the coefficients of a multi-loop
amplitude written as a linear combination of loop intergrals, as a
function of the momentum-twistor variables which can be evaluated,
either on $\Q$ or on finite fields $\Z_p$, via a sequence of
elementary rational operations.  The application of functional
reconstruction algorithms based on finite-field evaluations is thus
possible, and it can be an efficient way of computing scattering
amplitudes in gauge theories.  In the following we give more details
about how to combine these ingredients for tree-level calculations via
recursion relations, and for multi-loop calculations via integrand
reduction and generalized unitarity.

Similar concepts apply to the spinor-helicity formalism in higher
numbers of dimensions which, as stated, can be useful in order to
provide a higher-dimensional embedding of loop momenta.  Details on
the six-dimensional spinor-helicity formalism are given in
Appendix~\ref{sec:six-dimens-momenta}.

\subsection{Berends-Giele recursion}
\label{sec:berends-giele-recurs}
Berends-Giele recursion~\cite{Berends:1987me} is an efficient method
for the numerical calculation of tree-level amplitudes.  Even though
it is typically used with floating-point arithmetics, it can similarly
be applied with trivial modifications to finite fields $\Z_p$.

\begin{figure}[t]
  \centering
  \includegraphics[width=\textwidth]{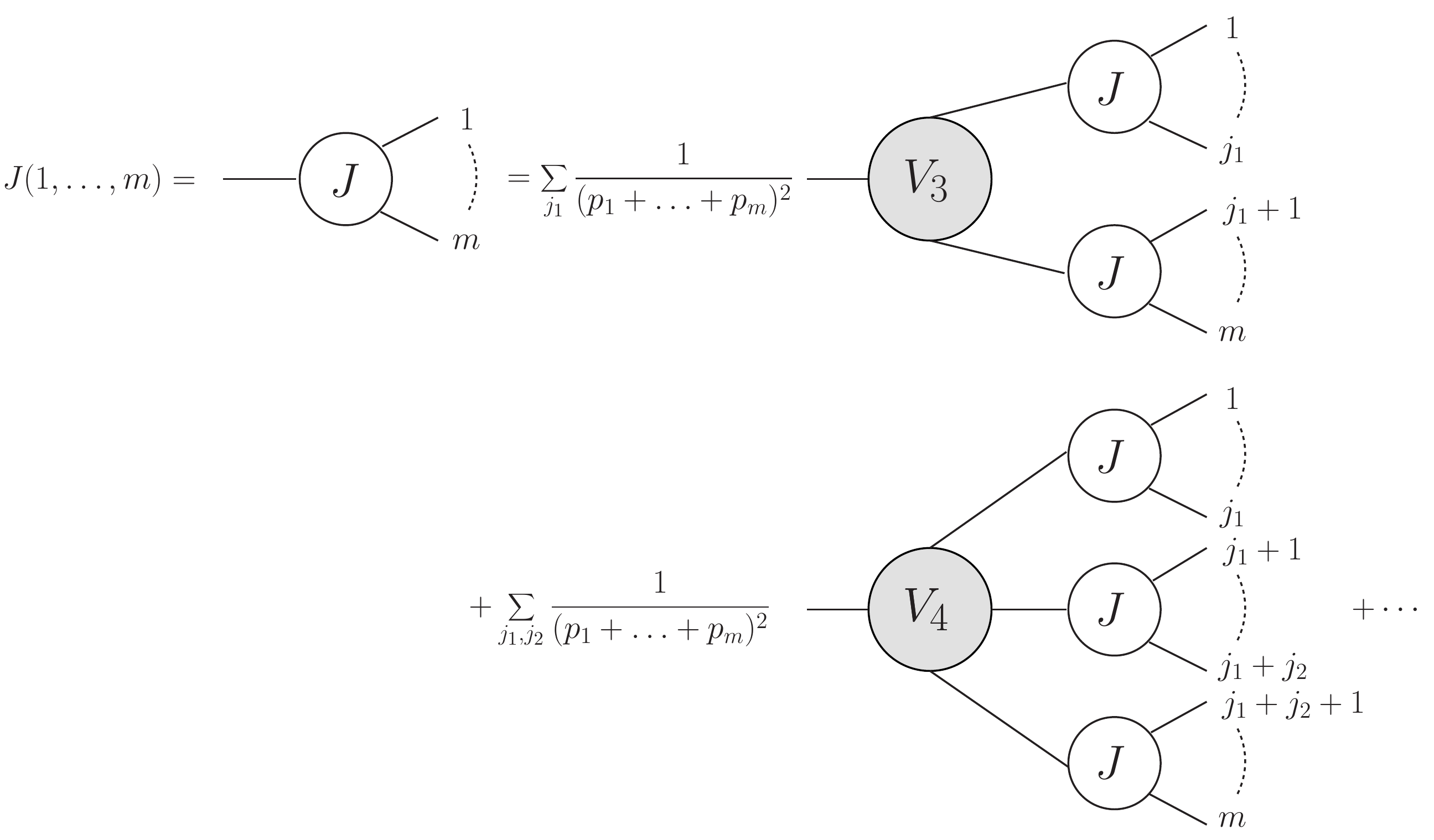}
  \caption{Schematic depiction of Berends-Giele recursion relations.
    The grey blobs $V_k$ represent contractions defined by the
    $k$-point Feynman rules of the theory.}
  \label{fig:bgrec}
\end{figure}

It is straightforward to apply the concepts outlined in
sect.~\ref{sec:four-dimens-momenta} for the four-dimensional case and
in Appendix~\ref{sec:six-dimens-momenta} for the six-dimensional one,
to Berends-Giele recursion.  One can start defining one-point on-shell
currents, which we symbolically denote by
\begin{equation}
  \includegraphics[scale=0.65, trim=0 8 0 0]{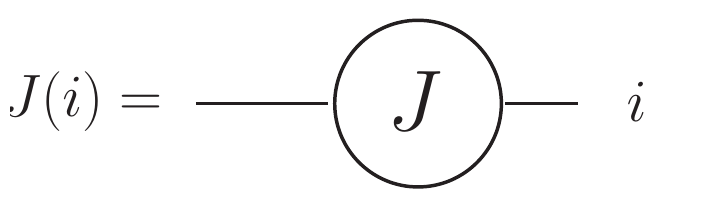},
\end{equation}
as external polarization vectors
($J(i)\equiv \epsilon_{h_i}^\mu(p_i)$), spinors
($J(i)\equiv|i\ra, |i ]$) or constants ($J(i)\equiv1$) depending on
the kind of particle involved, i.e.\ vector bosons, fermions, and
scalars respectively.  Then, higher-point off-shell currents are
computed by contracting two or more lower-point currents with
appropriate spinor and tensor structures which can be easily worked
out from the Feynman rules of the theory.  For color-ordered
amplitudes, the recursion for the calculation of an off-shell current
$J(1,\ldots,m)$ is depicted in Fig.~\ref{fig:bgrec}.
In the last step of the recursion for the calculation of an $n$-point
amplitude, an on-shell current $J(1,\ldots,n-1)$ is computed via a
similar recursion relation as the one in Fig.~\ref{fig:bgrec}, except
that there is no multiplication by the propagator factor, and then it
is contracted with the on-shell one-point current defined by the
$n$-th external leg.

Besides its efficiency, largely due to the fact that lower-point
currents can be re-used for the calculation of several higher-point
currents, Berends-Giele recursion also has the advantage that it can
be straightforwardly implemented for any theory and that the algorithm
is largely independent of the number of space-time dimensions.  For
these reasons Berends-Giele currents can also be efficiently used as
building blocks for multi-loop integrands, as we will describe in the
next section and in Appendix~\ref{sec:two-loop-unitarity}.

\section{Multi-loop integrand reduction and generalized unitarity}
\label{sec:multi-loop-integrand}

In this section we discuss a finite-field implementation of a specific
algorithm for the calculation of scattering amplitudes in gauge
theories, namely integrand reduction via $d$-dimensional generalized
unitarity.  The algorithm uses as building blocks tree-level
amplitudes and Berends-Giele currents, which in turn can be computed
using the spinor-helicity formalism as discussed in
sect.~\ref{sec:spinor-helicity-tree}.

We consider $\ell$-loop contributions to $n$-point scattering
amplitudes in dimensional regularization.  In particular we choose a
regularization scheme such that~\cite{Bern:2002zk}
\begin{itemize}
\item external momenta and polarizations are in four dimensions
\item loop momenta are in $d$ dimensions, with $d>4$
\item internal gluon states live in a $d_s$-dimensional space, with
  $d_s>d$.
\end{itemize}
The popular t'Hooft-Veltman\cite{tHooft:1972fi} and
Four-Dimensional-Helicity~\cite{Bern:2002zk} schemes can be obtained
as special cases, by setting $d_s=d$ and $d_s=4$ respectively at the
end of the calculation.

A generic contribution to a loop amplitude takes the form
\begin{equation} \label{eq:genericlloopint}
  \int_{-\infty}^{\infty} \left(\prod_{i=1}^\ell d^d k_i\right) \frac{\N(k_i)}{\prod_j D_j(k_i)},
\end{equation}
where $\N$ and $D$ are polynomials in the components of the loop
momenta $k_i$ (a rational dependence on the external kinematic
variables is always understood).  In particular, the denominators
$D_i$ correspond to loop propagators and have the generic quadratic
form
\begin{equation}
  D_i = \ell_i^2 - m_i^2,\qquad l_i^\mu = \sum_{j=1}^\ell \alpha_{ij} k_j^\mu + \sum_{j=1}^n \beta_{ij} p_j^\mu \quad (\alpha_{ij},\beta_{ij}\in \{0,\pm 1\}).
\end{equation}
It is often useful to split the loop momenta $k_i$ in a
four-dimensional part $k_i^{[4]}$ and a $(d-4)$-dimensional part
$k_i^{[d-4]}$ as
\begin{equation}
  k_i^\mu = k_i^{[4]}{}^\mu + k_i^{[d-4]}{}^\mu,
\end{equation}
and define extra-dimensional scalar products
\begin{equation} \label{eq:muij}
  \mu_{ij} = - \left( k_i^{[d-4]}\cdot k_j^{[d-4]} \right).
\end{equation}
In the regularization scheme defined above, a loop integrand can only
depend on the additional $(d-4)$-dimensional components of the loop
momenta through the scalar products $\mu_{ij}$ defined above.  In
particular, for one-loop amplitudes we only have one extra-dimensional
scalar product, i.e.\ $\mu_{11}$, while at two loops we have three of
them, namely $\mu_{11}$, $\mu_{22}$ and $\mu_{12}$.  The dependence on
these extra-dimensional components can be implemented by embedding the
loop momenta in a $\D$-dimensional space, where $\D$ is a sufficiently
large integer.  In particular, the choice $\D=6$ is sufficient for
both one- and two-loop applications.  An advantage of this
higher-dimensional embedding, especially in the context of generalized
unitarity, is that one can use a higher-dimensional spinor-helicity
formalism for the calculation of the integrands, having thus an
explicit representation of both external and internal states.  This
strategy has been used e.g.\ in
references~\cite{Badger:2013gxa,Badger:2016ozq}.  Having an explicit
representation of loop momenta and spinors is also obviously
advantageous in numerical calculations, including computations on
finite fields $\Z_p$, which make this formalism suited for our goals.

Integrand reduction methods rewrite loop integrands of the form of
Eq.~\eqref{eq:genericlloopint} as a sum of irreducible contributions,
\begin{equation}
  \frac{\N(k_i)}{\prod_j D_j(k_i)} = \sum_{T} \frac{\Delta_T(k_i)}{\prod_{j\in T} D_j(k_i)},
\end{equation}
where the sum on the r.h.s.\ runs over the sub-topologies $T$ of the
parent topology identified by the set of denominators $D_j$.  The
so-called on-shell numerators $\Delta_T$, also known in this context
as residues, are linear combinations of basis elements
$\{\m_T^\alpha\}$, namely
\begin{equation} \label{eq:deltaT}
  \Delta_{T}(k_i) = \sum_{\alpha} c_{T,\alpha}\, \left(\m_T(k_i)\right)^\alpha,
\end{equation}
where $\alpha$ runs over an appropriate topology-dependent set of
multi-indexes and $\m_T$ represents a sequence of polynomials in the
loop momenta.  The coefficients $c_{T,\alpha}$, which only depend on
the external kinematics, can be obtained by evaluating the integrand
on values of the loop momenta satisfying the so-called multiple-cut
conditions $\{D_j=0\}_{j\in T}$.  This corresponds to put on-shell a
subset of the loop momenta.  In particular, by evaluating the
integrand on several solutions of the cut constraints, one obtains a
linear system of equations for the coefficients $c_{T,\alpha}$ which
can be solved e.g.\ via Gaussian elimination.

In general, there is no unique choice for the set of basis elements
$\{\m_T^\alpha\}$, which are only constrained by the requirements of
being independent of the aforementioned cut conditions and forming a
complete integrand basis compatible with the rank of numerator with
respect to the loop momenta, up to terms proportional to the
denominators $D_i$ of the topology $T$.  Techniques for choosing an
appropriate integrand basis have been proposed
elsewhere~\cite{Badger:2012dp,Zhang:2012ce,Mastrolia:2012an,Badger:2016ozq}
and their discussion is outside the purposes of this paper.  Here we
will limit ourself to specify case-by-case the integrand basis we used
in each example.

As stated, in the calculations presented in this paper, we embed the
loop momenta in a $\D$-dimensional space (with $\D=6$).  Each loop
momentum $k_i$ is thus decomposed into a basis $\{e_{ij}\}_{j=1}^\D$,
\begin{equation}
  k_i^\mu = \sum_{j=1}^\D y_{ij}\, e_{ij}^\mu.
\end{equation}
For each cut, we identify a set of independent free parameters
$\{\tau_k\}$ which describe the set of solutions.  In particular we
look for a set of variables such that the coefficients
$y_{ij}=y_{ij}(\tau_k)$ of the linear combination are rational
functions of the parameters $\tau_k$, which is particularly convenient
when working with finite fields.\footnote{ Irrational solutions might
  also be accommodated by considering fields that are more general
  than $\Z_p$, at the price of making some intermediate steps of the
  calculation more involved.}  We point out that this has been shown
to be possible in many two-loop examples in $d$
dimensions~\cite{Badger:2013gxa,Mastrolia:2016dhn,Badger:2016ozq},
while in general it is not in the four-dimensional limit.  Using
$\D$-dimensional cut loop momenta defined by assigning numerical
values to the free parameters $\tau_k$, we thus evaluate both the
integrand the on-shell basis $\m_T^\alpha$, generating the linear
system of equations we can solve for the coefficients $c_{T,\alpha}$
in Eq.~\eqref{eq:deltaT}.

The evaluation of the integrand on multiple cuts might be done by
means of its analytic expression, if available (e.g.\ via diagrammatic
techniques).  The method of generalized unitarity instead exploits the
fact that a multi-loop integrand, when a subset of the loop momenta
are put on-shell by the cut constraints, factorizes as a product of
tree-level amplitudes.  This factorization can be understood by the
fact that the numerator of an on-shell propagator factorizes as a sum
of polarization states.  Using generalized unitarity, the amplitude is
expressed in terms of a smaller number of contributions compared with
diagrammatic techniques, and gauge invariance is guaranteed in
intermediate steps of the calculation, thus avoiding cancellations of
gauge-dependent terms between different diagrams.  Moreover, it offers
the possibility of exploiting efficient tree-level techniques for loop
calculations as well.  When using Berends-Giele currents, additional
simplifications are possible during the evaluation of the cut
integrands, along the lines of what is implemented in the one-loop
public code \textsc{NJet}~\cite{Badger:2012pg}, whose two-loop
extension is briefly discussed in
Appendix~\ref{sec:two-loop-unitarity}.

\begin{figure}[t]
  \centering
  \includegraphics[width=\textwidth]{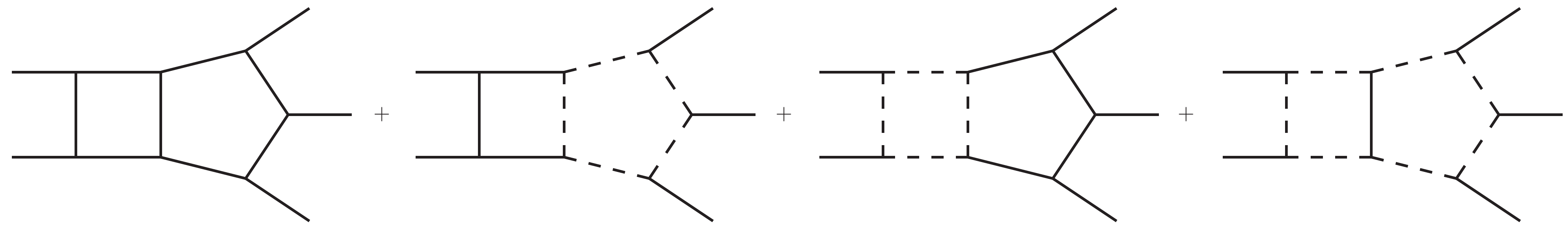}
  \caption{Sum of diagrams with gluon (solid lines) and scalar (dashed
    lines) loops, for the penta-box topology.  Each scalar loop should
    be multiplied by the number of scalar flavours, which in our case
    is equal to $d_s-\D$.}
  \label{fig:pentaboxssum}
\end{figure}

Since we embed loop momenta in $\D=6$ dimensions, we can evaluate the
amplitudes relevant for each cut using the six-dimensional
spinor-helicity techniques presented in ref.~\cite{Cheung:2009dc} and
whose application to finite-field calculations is discussed in
Appendix~\ref{sec:six-dimens-momenta}.  We also add to the theory
$d_s-\D$ flavours of scalars representing additional polarizations of
the internal gluons which, as stated at the beginning of this section,
are taken to be $d_s$-dimensional.  In particular, each on-shell
integrand at two-loops can be written as
\begin{equation}
  \Delta_T = \Delta_T^{(\D,0)} + (d_s-\D)\, \Delta_T^{(\D,1)} + (d_s-\D)^2\, \Delta_T^{(\D,2)},
\end{equation}
where $\Delta_T^{(\D,i)}$ is a $\D$-dimensional integrand with $i$
scalar loops.  Notice that the result for $\Delta_T$ does not depend
on the dimension $\D$ of the chosen embedding, unlike each of the
terms on the r.h.s.\ of the previous equation.  Moreover, for pure
Yang-Mills theories, the
contribution $\Delta_T^{(\D,2)}$ is non-vanishing only for one-loop
squared topologies.  As an example, the two-loop planar penta-box is
obtained as the sum of contributions in fig.~\ref{fig:pentaboxssum}.
This is the same set-up used for the analytic calculations presented
in references~\cite{Badger:2013gxa,Badger:2015lda,Badger:2016ozq}.

In the following examples, we apply the functional reconstruction
algorithm to the coefficients $c_{T,\alpha}^{(i)}$, with $i=0,1,2$,
defined by the expansion of each coefficient $c_{T,\alpha}$ of
Eq.~\eqref{eq:deltaT} in powers of $d_s-2$,
\begin{equation}
  c_{T,\alpha} = c_{T,\alpha}^{(0)}+(d_s-2)\, c_{T,\alpha}^{(1)}+(d_s-2)^2\, c_{T,\alpha}^{(2)}.
\end{equation}
Using $d_s-2$, rather than just $d_s$, is motivated by the fact that
$d_s-2$ is the number of internal gluon polarizations, and indeed this
choice makes the coefficients $c_{T,\alpha}^{(i)}$ significantly
simpler.  In our implementation, because a numerical evaluation
consists in solving for all the coefficients $c_{T,\alpha}^{(i)}$ at
the same time, we cache their results and reuse them across the
functional reconstruction of all the coefficients.  Because the
examples below involve genuine two-loop topologies (as opposed to
one-loop squared topologies), in these cases we always have
$c_{T,\alpha}^{(2)}=0$.  All the analytic results are publicly
available, and should be useful for comparisons with future
calculations.\footnote{ They can be obtained at the url
  \texttt{https://bitbucket.org/peraro/ff2lexamples}.}

\subsection{Two-loop five-point planar penta-box}
\label{sec:2lpb}

We consider the two-loop penta-box topology, where the five external
gluons are taken as outgoing.
\begin{figure}[t]
  \centering
  \includegraphics[width=0.5\textwidth]{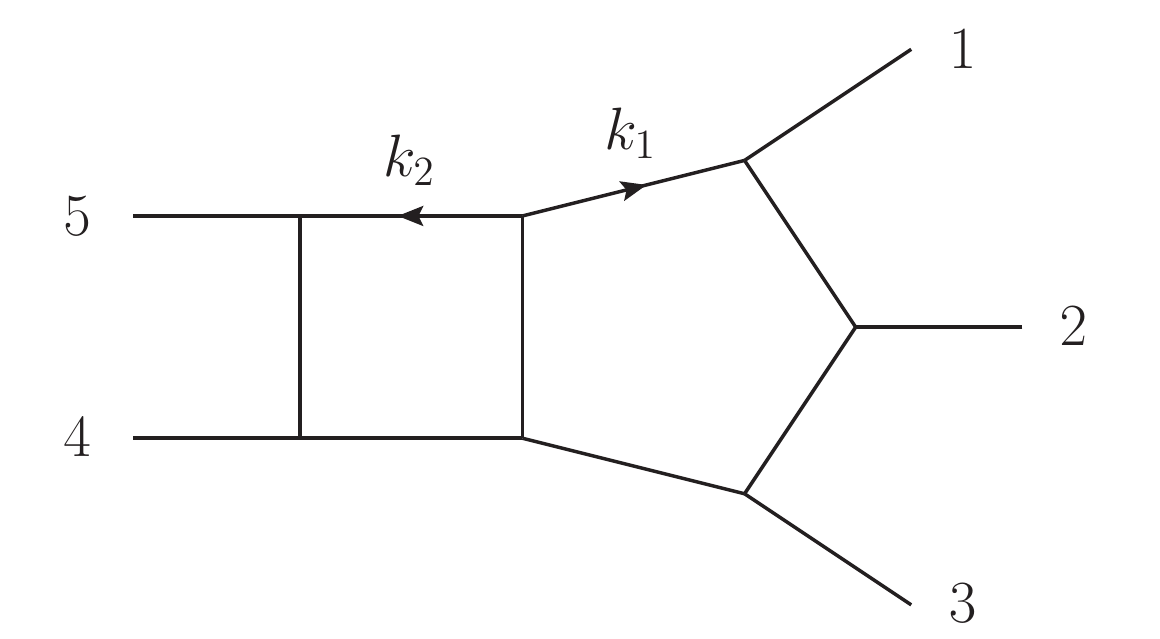}
  \caption{The two-loop planar penta-box topology.}
  \label{fig:pentabox}
\end{figure}
The loop momenta $k_1$ and $k_2$ are defined as in
fig.~\ref{fig:pentabox}.

As stated, the kinematics is defined by the spinor components, which
in turn are parametrized by momentum-twistor variables.  For the
five-point case we use the parametrization given
in~\cite{Badger:2013gxa}, namely
\begin{alignat}{2} \label{eq:5pttw}
  |1\ra =& {} {1 \choose 0}, \qquad & |1]  ={}&  {1 \choose \frac{x_4-x_5}{x_4}}, \nn
  |2\ra =& {} {0 \choose 1}, \qquad & |2]  ={}&  {0 \choose x_1}, \nn
  |3\ra =&{} {\frac{1}{x_1} \choose 1}, \qquad& |3]  ={}&  {x_1\, x_4 \choose - x_1}, \nn
  |4\ra =& {} {\frac{1}{x_1}+\frac{1}{x_1\, x_2} \choose 1}, & \qquad |4]  ={}&  {x_1(x_2\, x_3-x_3\, x_4-x_4) \choose -\frac{x_1\, x_2\, x_3\, x_5}{x_4}}, \nn
  |5\ra =& {} {\frac{1}{x_1}+\frac{1}{x_1\, x_2}+\frac{1}{x_1\, x_2\, x_3} \choose 1}, & \qquad |5]  ={}&  {x_1\, x_3 (x_4-x_2) \choose \frac{x_1\, x_2\, x_3\, x_5}{x_4}}.
\end{alignat}
The variables $x_i$ can be expressed in terms of the Mandelstam
invariants $s_{ij}$ and $\text{tr}_5(1\, 2\, 3\, 4)$ using Eq.~(A.8)
of ref.~\cite{Badger:2013gxa}.  In particular, all the
momentum-twistor variables but $x_1$ (we recall that $x_1=s_{12}$) are
dimensionless.  This means that the dependence of the result on $x_1$
can be fixed by dimensional analysis and the functional reconstruction
will thus use the following set of variables
\begin{equation}
  \z = (x_2,x_3,x_4,x_5).
\end{equation}

As discussed in sect.~\ref{sec:four-dimens-momenta}, we factor out of
the result a global helicity-dependent phase $\A^{(\text{phase})}$.
In this example we define it as
\begin{align} \label{eq:5ptphase}
  \A^{(\text{phase})}(1^+,2^+,3^+,4^+,5^+) ={}& i\, \frac{s_{12}^6}{\la 1\, 2 \ra\, \la 2\, 3 \ra\, \la 3\, 4 \ra\, \la 4\, 5 \ra\, \la 5\, 1 \ra} \nn
  \A^{(\text{phase})}(1^-,2^+,3^+,4^+,5^+) ={}& i\, \frac{\left(\la 1\, 2\ra\, [ 2 3 ] \la 3\, 1\ra]\right)^2 s_{12}^3}{\la 1\, 2 \ra\, \la 2\, 3 \ra\, \la 3\, 4 \ra\, \la 4\, 5 \ra\, \la 5\, 1 \ra} \nn
  \A^{(\text{phase})}(1^-,2^-,3^+,4^+,5^+) ={}& i\, \frac{\la 1\, 2\ra^4 s_{12}^4}{\la 1\, 2 \ra\, \la 2\, 3 \ra\, \la 3\, 4 \ra\, \la 4\, 5 \ra\, \la 5\, 1 \ra},\nn
  \A^{(\text{phase})}(1^-,2^+,3^-,4^+,5^+) ={}& i\, \frac{\la 1\, 3\ra^4 s_{12}^4}{\la 1\, 2 \ra\, \la 2\, 3 \ra\, \la 3\, 4 \ra\, \la 4\, 5 \ra\, \la 5\, 1 \ra},
\end{align}
and similar for all the ones obtained by cyclic permutations of the
external legs.  Helicity configurations with three or more negative
helicities are not listed, since they can be obtained from the ones
above and their cyclic permutations by charge conjugation.

Notice that charge conjugation corresponds to swapping the spinors
$|i\ra \leftrightarrow |i]$ both in the definition of
$\A^{(\text{phase})}$ and in their parametrization in terms of
momentum-twistor variables.  In this case the definition of the $x_i$
is modified by applying the same transformation, which amounts to the
substitution $\text{tr}_5\to -\text{tr}_5$ in Eq.~(A.8) of
ref.~\cite{Badger:2013gxa}.

Given the symmetries of the penta-box diagram, all the helicity
configurations can be obtained from the following set
\begin{align}
  \{&(1^+,2^+,3^+,4^+,5^+),\nn
    &(1^-,2^+,3^+,4^+,5^+),\ (1^+,2^-,3^+,4^+,5^+),\ (1^+,2^+,3^+,4^-,5^+),\nn
    &(1^-,2^-,3^+,4^+,5^+),\ (1^-,2^+,3^+,4^+,5^-),\ (1^+,2^+,3^+,4^-,5^-),\nn
    &(1^-,2^+,3^-,4^+,5^+),\ (1^+,2^-,3^+,4^-,5^+),\ (1^-,2^+,3^+,4^-,5^+)\}.
\end{align}
For each of these, the on-shell integrand is parametrized as
\begin{equation}
  \Delta_{\textrm{pb}} = \A^{(\text{phase})}\, \sum_{\alpha}\, (s_{12})^{-|\alpha|}\, c_{\textrm{pb}, \alpha}\,  \m_{\textrm{pb}}^\alpha,
\end{equation}
where
\begin{equation}
  c_{\textrm{pb}, \alpha} = c_{\textrm{pb}, \alpha}^{(0)}(x_2,x_3,x_4,x_5) + (d_s-2)\, c_{\textrm{pb}, \alpha}^{(1)}(x_2,x_3,x_4,x_5),
\end{equation}
and the $c_{\textrm{pb}, \alpha}^{(i)}$ are computed via the
functional reconstruction method previously illustrated in this paper.
The integrand basis is chosen as in ref.~\cite{Badger:2013gxa}, namely
\begin{equation}
  \m_{\textrm{pb}} = (2\, (k_1\cdot p_5),\  2\, (k_2\cdot p_2),\  2\, (k_2\cdot p_1),\ \mu_{11},\ \mu_{12},\  \mu_{22}),
\end{equation}
with
\begin{align}
  \alpha \in \{ & (0, 0, 0, 0, 0, 0),
       (0, 0, 0, 0, 0, 1),
       (0, 0, 0, 0, 1, 0),
       (0, 0, 0, 1, 0, 0),
       (0, 0, 1, 0, 0, 0),\nn &
       (0, 1, 0, 0, 0, 0),
       (1, 0, 0, 0, 0, 0),
       (0, 0, 0, 0, 0, 2),
       (0, 0, 0, 0, 1, 1),
       (0, 0, 0, 1, 0, 1),\nn &
       (0, 0, 1, 0, 0, 1),
       (0, 1, 0, 0, 0, 1),
       (1, 0, 0, 0, 0, 1),
       (0, 0, 0, 0, 2, 0),
       (0, 0, 0, 1, 1, 0),\nn &
       (0, 0, 1, 0, 1, 0),
       (0, 1, 0, 0, 1, 0),
       (1, 0, 0, 0, 1, 0),
       (0, 0, 0, 2, 0, 0),
       (0, 0, 1, 1, 0, 0),\nn &
       (0, 1, 0, 1, 0, 0),
       (1, 0, 0, 1, 0, 0),
       (0, 0, 0, 1, 0, 2),
       (1, 0, 0, 0, 0, 2),
       (0, 0, 0, 0, 2, 1),\nn &
       (0, 0, 0, 1, 1, 1),
       (0, 0, 1, 0, 1, 1),
       (0, 1, 0, 0, 1, 1),
       (1, 0, 0, 0, 1, 1),
       (0, 0, 0, 2, 0, 1),\nn &
       (0, 0, 1, 1, 0, 1),
       (0, 1, 0, 1, 0, 1),
       (1, 0, 0, 1, 0, 1),
       (0, 0, 2, 0, 0, 1),
       (0, 1, 1, 0, 0, 1),\nn &
       (0, 0, 0, 0, 3, 0),
       (0, 0, 0, 1, 2, 0),
       (0, 0, 1, 0, 2, 0),
       (0, 1, 0, 0, 2, 0),
       (1, 0, 0, 0, 2, 0),\nn &
       (0, 0, 0, 2, 1, 0),
       (0, 0, 1, 1, 1, 0),
       (0, 1, 0, 1, 1, 0),
       (1, 0, 0, 1, 1, 0),
       (0, 0, 2, 0, 1, 0),\nn &
       (0, 1, 1, 0, 1, 0),
       (0, 0, 0, 3, 0, 0),
       (0, 0, 1, 2, 0, 0),
       (0, 1, 0, 2, 0, 0),
       (1, 0, 0, 2, 0, 0),\nn &
       (0, 0, 2, 1, 0, 0),
       (0, 1, 1, 1, 0, 0),
       (1, 0, 0, 1, 0, 2),
       (1, 0, 0, 0, 2, 1),
       (0, 0, 1, 1, 1, 1),\nn &
       (0, 1, 0, 1, 1, 1),
       (1, 0, 0, 1, 1, 1),
       (0, 0, 1, 2, 0, 1),
       (0, 1, 0, 2, 0, 1),
       (1, 0, 0, 2, 0, 1),\nn &
       (0, 0, 1, 0, 3, 0),
       (0, 1, 0, 0, 3, 0),
       (1, 0, 0, 0, 3, 0),
       (0, 0, 1, 1, 2, 0),
       (0, 1, 0, 1, 2, 0),\nn &
       (1, 0, 0, 1, 2, 0),
       (0, 0, 2, 0, 2, 0),
       (0, 1, 1, 0, 2, 0),
       (0, 0, 1, 2, 1, 0),
       (0, 1, 0, 2, 1, 0),\nn &
       (1, 0, 0, 2, 1, 0),
       (0, 0, 3, 0, 1, 0),
       (0, 1, 2, 0, 1, 0),
       (0, 0, 1, 3, 0, 0),
       (0, 1, 0, 3, 0, 0),\nn &
       (0, 0, 3, 1, 0, 0),
       (0, 1, 2, 1, 0, 0),
       (0, 0, 4, 1, 0, 0),
       (0, 1, 3, 1, 0, 0)\}.
\end{align}
This basis is smooth in the four-dimensional limit $\mu_{ij}\to 0$.

The results we obtain for the all-plus topology agree with those
computed in references~\cite{Badger:2013gxa,Badger:2016ozq}.  The
on-shell integrands for the other helicity configurations are 
new.  The results are summarized in table~\ref{tab:pentabox}.
\begin{table}[t]
  \centering
  \begin{tabular}{| l | r | r | r | r |}
    \hline
    Helicity & Non-vanishing coeff. & Max.\ terms & Max.\ degree & Avg.\ non-zero terms \\
    \hline
    \hline
    $(1^+,2^+,3^+,4^+,5^+)$ & 14 & 19 & 8 & 15.00 \\
    $(1^-,2^+,3^+,4^+,5^+)$ & 27 & 443 & 19 & 152.96 \\
    $(1^+,2^-,3^+,4^+,5^+)$ & 37 & 1977 & 24 & 674.97 \\
    $(1^+,2^+,3^+,4^-,5^+)$ & 61 & 474 & 18 & 184.05 \\
    $(1^-,2^-,3^+,4^+,5^+)$ & 35 & 1511 & 24 & 278.77 \\
    $(1^-,2^+,3^+,4^+,5^-)$ & 79 & 7027 & 34 & 1112.82 \\
    $(1^+,2^+,3^+,4^-,5^-)$ & 18 & 19 & 8 & 15.00 \\
    $(1^-,2^+,3^-,4^+,5^+)$ & 41 & 2412 & 22 & 368.41 \\
    $(1^+,2^-,3^+,4^-,5^+)$ & 85 & 18960 & 42 & 3934.96 \\
    $(1^-,2^+,3^+,4^-,5^+)$ & 85 & 10386 & 37 & 1803.52 \\
    \hline
  \end{tabular}
  \caption{Summary of the results for the planar penta-box topology.
    For each helicity configuration we list the number of
    non-vanishing coefficients $c_{\textrm{pb},\alpha}^{(i)}$, the
    maximum number of terms in a coefficient, the maximum degree (of
    either the numerator or the denominator) of a coefficient, and the
    average number of terms in the non-vanishing coefficients.  The
    number of terms in a rational function is defined as the sum of the number of
    monomials in its numerator and the one in its denominator.  Full
    analytic expressions are available at the url
    \texttt{https://bitbucket.org/peraro/ff2lexamples}.}
  \label{tab:pentabox}
\end{table}

\subsection{Two-loop five-point non-planar double-pentagon}
\label{sec:2ldp}
\begin{figure}[t]
  \centering
  \includegraphics[width=0.5\textwidth]{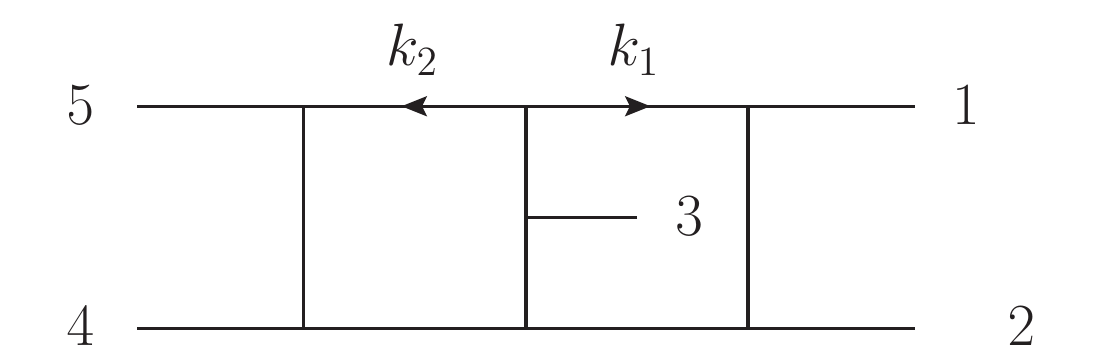}
  \caption{The two-loop non-planar double-pentagon topology.}
  \label{fig:doublepentagon}
\end{figure}

We now consider the two-loop non-planar double-pentagon topology
depicted in fig.~\ref{fig:doublepentagon}.
The set-up of the calculation is completely analogous to the one for
the penta-box.  The kinematics and the global spinor phase are still
defined using Eq.~\eqref{eq:5pttw} and Eq.~\eqref{eq:5ptphase}
respectively.

For this topology, we consider the following complete set of
independent helicity configurations
\begin{align}
  \{&(1^+,2^+,3^+,4^+,5^+),\nn
    &(1^-,2^+,3^+,4^+,5^+),\ (1^+,2^+,3^-,4^+,5^+),\nn
    &(1^-,2^-,3^+,4^+,5^+),\ (1^+,2^-,3^-,4^+,5^+),\nn
    &(1^-,2^+,3^+,4^+,5^-),\ (1^-,2^+,3^+,4^-,5^+)\}.
\end{align}

For this example we use a slightly different approach for the
parametrization of the integrand.  Rather than using an integrand
basis with a smooth four-dimensional limit $\mu_{ij}\to 0$, we choose
one which is only composed of scalar products between loop momenta and
external momenta.  While this, in general, might make some of the
coefficients more complex and the integration harder, it has the
advantage that the result can be directly plugged into available
programs for integration by parts and other multi-loop techniques.
This is also consistent with the adaptive integrand decomposition
recently proposed in ref.~\cite{Mastrolia:2016dhn}, where the
integrands are always expressed in terms of this kind of scalar
product and the four-dimensional components of the loop which are
orthogonal to the external legs are integrated out (as this is a
five-point topology, it corresponds to a limiting case where no such
orthogonal direction is present).  More in detail, we use
\begin{equation}
  \Delta_{\textrm{dp}} = \A^{(\text{phase})}\, \sum_{\alpha}\, (s_{12})^{-|\alpha|}\, c_{\textrm{dp}, \alpha}\,  \m_{\textrm{dp}}^\alpha,
\end{equation}
where, as before,
\begin{equation}
  c_{\textrm{dp}, \alpha} = c_{\textrm{dp}, \alpha}^{(0)}(x_2,x_3,x_4,x_5) + (d_s-2)\, c_{\textrm{dp}, \alpha}^{(1)}(x_2,x_3,x_4,x_5),
\end{equation}
while the integrand basis is now given by
\begin{equation}
  \m_{\textrm{dp}} = (2\, (k_1\cdot p_5),\  2\, (k_2\cdot p_1),\  2\, (k_1\cdot p_3)-2\, (k_2\cdot p_3)),
\end{equation}
with
\begin{align}
  \alpha \in \{& (0, 0, 0),
        (0, 0, 1),
        (0, 0, 2),
        (0, 0, 3),
        (0, 0, 4),
        (0, 0, 5),
        (0, 1, 0),
        (0, 1, 1),
        (0, 1, 2),
        (0, 1, 3),\nn &
        (0, 1, 4),
        (0, 2, 0),
        (0, 2, 1),
        (0, 2, 2),
        (0, 2, 3),
        (0, 3, 0),
        (0, 3, 1),
        (0, 3, 2),
        (0, 4, 0),
        (0, 4, 1),\nn &
        (0, 5, 0),
        (1, 0, 0),
        (1, 0, 1),
        (1, 0, 2),
        (1, 0, 3),
        (1, 0, 4),
        (1, 1, 0),
        (1, 1, 1),
        (1, 1, 2),
        (1, 1, 3),\nn &
        (1, 1, 4),
        (1, 2, 0),
        (1, 2, 1),
        (1, 2, 2),
        (1, 2, 3),
        (1, 3, 0),
        (1, 3, 1),
        (1, 3, 2),
        (1, 4, 0),
        (1, 4, 1),\nn &
        (1, 5, 0),
        (2, 0, 0),
        (2, 0, 1),
        (2, 0, 2),
        (2, 0, 3),
        (2, 1, 0),
        (2, 1, 1),
        (2, 1, 2),
        (2, 1, 3),
        (2, 2, 0),\nn &
        (2, 2, 1),
        (2, 2, 2),
        (2, 2, 3),
        (2, 3, 0),
        (2, 3, 1),
        (2, 3, 2),
        (2, 4, 0),
        (2, 4, 1),
        (2, 5, 0),
        (3, 0, 0),\nn &
        (3, 0, 1),
        (3, 0, 2),
        (3, 1, 0),
        (3, 1, 1),
        (3, 1, 2),
        (3, 2, 0),
        (3, 2, 1),
        (3, 2, 2),
        (3, 3, 0),
        (3, 3, 1),\nn &
        (3, 4, 0),
        (4, 0, 0),
        (4, 0, 1),
        (4, 1, 0),
        (4, 1, 1),
        (4, 2, 0),
        (4, 2, 1),
        (4, 3, 0),
        (5, 0, 0),
        (5, 1, 0),\nn &
        (5, 2, 0)\}.
\end{align}

The results we obtain for the all-plus topology are in agreement with
those computed in ref.~\cite{Badger:2015lda}, although in that
reference they were deduced from the known expressions in the planar
case, while here they have been directly computed using
integrand-reduction via generalized unitarity on the non-planar
topology.  The on-shell integrands for the other helicity
configurations are computed here for the first time.  The results are
summarized in table~\ref{tab:doublepentagon}.
\begin{table}[t]
  \centering
  \begin{tabular}{| l | r | r | r | r |}
    \hline
    Helicity & Non-vanishing coeff. & Max.\ terms & Max.\ degree & Avg.\ non-zero terms \\
    \hline
    \hline
    $(1^+,2^+,3^+,4^+,5^+)$ & 104 & 1937 & 26 & 626.39 \\
    $(1^-,2^+,3^+,4^+,5^+)$ & 104 & 1449 & 27 & 601.43 \\
    $(1^+,2^+,3^-,4^+,5^+)$ & 104 & 1554 & 23 & 642.90 \\
    $(1^-,2^-,3^+,4^+,5^+)$ & 99 & 1751 & 26 & 739.05 \\
    $(1^+,2^-,3^-,4^+,5^+)$ & 104 & 2524 & 24 & 923.71 \\
    $(1^-,2^+,3^+,4^+,5^-)$ & 104 & 1838 & 27 & 823.00 \\
    $(1^-,2^+,3^+,4^-,5^+)$ & 104 & 1307 & 24 & 630.48 \\
    \hline
  \end{tabular}
  \caption{Summary of the results for the non-planar double-pentagon
    topology.  The entries are analogous to those in
    table~\ref{tab:pentabox}.  As in the previous case, full analytic
    expressions are available at the url
    \texttt{https://bitbucket.org/peraro/ff2lexamples}.}
  \label{tab:doublepentagon}
\end{table}

\section{Implementation}
\label{sec:implementation}

In this section we give some detail about our implementation of the
functional reconstruction method illustrated in
sect.~\ref{sec:funct-reconstr} and its application to the techniques
described in sections~\ref{sec:spinor-helicity-tree}
and~\ref{sec:multi-loop-integrand}.

The programming language of our implementation is \textsc{C++}, and in
particular the \textsc{C++-11} standard of the language.\footnote{ We
  make extensive use of the \texttt{std::unique\_ptr} and
  \texttt{std::unordered\_map} containers, as well as right-value and
  move semantics, for the implementation of our data structures, in
  order to optimize performance.}

The first design choice regards the definition of the finite fields
$\Z_p$.  Our requirement of working with machine-size integers imposes
an upper limit $p<P_{\textrm{max}}$ on the choice of the primes $p$
defining the fields.  We work with 64-bit unsigned integers, which can
take values between $i_{\textrm{min}}=0$ and
$i_{\textrm{max}}=2^{64}-1$.  The most stringent requirement is
avoiding integer overflow in multiplication, i.e.\ the requirement
that by multiplying two integers in $\Z_p$ with $p<P_{\textrm{max}}$,
before taking modulo $p$ of the result, we still obtain a value
smaller than $i_{\textrm{max}}$.  This implies we can choose
$P_{\textrm{max}}$ such that
$P_{\textrm{max}}^2\leq i_{\textrm{max}}$.  In our implementation we
find more practical to use a more conservative choice, namely
$C\, P_{\textrm{max}}^3<i_{\textrm{max}}$ with $C=9$.  This allows us
to be slightly more relaxed in the implementation and take (sums of)
products of three integers in $\Z_p$ before taking modulo $p$.  This
approach also drastically reduces the number of $\mmod p$ operations
which are needed.  We hard-code 60 primes satisfying these criteria,
in the interval $897\, 473\leq p \leq 978\, 947$.
With this choice, in most of the examples presented in this paper we
only need one or two primes $p$ for the rational reconstruction of the
result over $\Q$ (according to the discussion in
sect.~\ref{sec:rati-reconstr} and Appendix~\ref{sec:mult-inverse}).

One important difference between native operations which involve
machine integers or floating-points numbers and operations on elements
of a finite field $\Z_p$ concerns division.  Indeed, as stated,
performing a division in $\Z_p$ requires the calculation of a
multiplicative inverse (see Appendix~\ref{sec:mult-inverse} for an
algorithm).  Hence, division in $\Z_p$ is a relatively expensive
operation (compared to other native operations) which needs a call
to a specific routine, and it is generally a good idea to minimize the
number of times it is required.  However, because the calculation of a
multiplicative inverse is implemented by a routine, catching divisions
by zero is very easy.  Notice that such cases can be quite common
during a functional reconstruction procedure, where many values for
the variables $\z$ are probed.  This makes detecting (actual or
spurious) singularities in the evaluation of a function relatively
easy and, as stated, the reconstruction algorithms illustrated above
can accomodate this possibility.

We now briefly describe our internal representation of polynomials and
rational functions.  For univariate Newton polynomials, defined as in
Eq.~\eqref{eq:unewtonrep}, we store two arrays of integers, namely the
coefficients $a_i$ and the values $y_i$.  A completely analogous
representation is used for Thiele's rational functions in
Eq.~\eqref{eq:thielerep}.  For multivariate Newton polynomials
depending on $n$ variables, recursively defined as in
Eq.~\eqref{eq:mnewtonrep}, we store the array of integers $y_i$ and an
array of pointers to the coefficients $a_i$, which are now Newton
polynomials in $n-1$ variables.  For our canonical representation of
polynomials, given in Eq.~\eqref{eq:defpoly}, we adopt instead a
sparse representation.  More in detail, we store an array of pointers
to a data structure representing each non-vanishing term.  These terms
are in turn stored as contiguous chunks of memory containing the
coefficient $c_\alpha$ and the exponent $\alpha$ corresponding to each
monomial.  We also store the total degree $|\alpha|$ for convenience.
The internal representation uses the graded lexicographic monomial
order (i.e.\ terms are ordered by their total degree, and terms with
the same total degree are ordered lexicographically).  We found this
representation suited for implementing the basic polynomial operations
needed by our reconstruction algorithm, namely polynomial addition,
multiplication by linear univariate monomials, and homogenization with
respect to one variable.  In particular the last two operations take
advantage of the possibility of manipulating the entries of the
exponents $\alpha$, which is trivial using our sparse representation.
Finally, rational functions are simply stored as a pair of polynomials
representing the numerator and the denominator respectively.

As stated, the only technique discussed in this paper which requires
multi-precision arithmetic is the rational reconstruction algorithm.
In our implementation we use the popular GNU Multiple Precision
Arithmetic Library (GMP).  The final result over the field $\Q$ is
stored using a sparse representation similar to the one we use for
$\Z_p$.  For functions over $\Q$ however we do not implement any of
the polynomial algorithms mentioned above, but only the routines
needed for merging results over several finite fields $\Z_{p_i}$ to
obtain a guess in $\Q$, as discussed in sect.~\ref{sec:rati-reconstr}
and Appendix~\ref{sec:chinese-remainder}.

It should be stressed that the results obtained with the illustrated
reconstruction algorithm for rational functions is minimal with
respect to the total degree of the numerator and the denominator,
hence there is no need to import it into an external computer algebra
system for polynomial GCD simplification (although it can obviously
be used by algebra systems for other kinds of manipulations).

It is worth making a few remarks about the application of the
reconstruction algorithm to the examples described in
sect.~\ref{sec:multi-loop-integrand}.  As stated, the coefficients
$c_{T,\alpha}^{(i)}$ to be reconstructed for each topology $T$ are
computed by evaluating the loop integrands on the multiple cuts and
thus obtaining a linear system of equations.  The latter can be solved
for them by means of Gauss elimination, for which we have a very
straightforward implementation suited for dense systems.  In order to
minimize the number of integrand evaluations we need, as well as the
size of the system of equations, we first evaluate each coefficient
for a set of arbitrary values of the variables $\z$ and the primes $p$
defining the field $\Z_p$.  These evaluations are used to collect
information about which coefficients are non-vanishing for each
helicity configuration.  Then, when applying the functional
reconstruction algorithm, we restrict the system to the non-vanishing
coefficients only.  With fewer unknowns, we thus require fewer
evaluations of the integrand in order to find a solution.  Moreover,
as stated, because the solution of the system returns all the
non-vanishing coefficients, we cache their value for each $\z$ in a
hash table.  This allows to quickly look up the values of
$c_{T,\alpha}^{(i)}$ if they have already been computed during the
reconstruction of other coefficients.

Our current implementation can be considered a proof-of-concept and
lacks many optimizations which can be estimated to improve the
performance by at least one order of magnitude.  However it is already
suited for applications to complex problems in high-energy physics
such as multi-loop high-multiplicity calculations, where it can easily
outperform equivalent fully-analytic approaches.  In the examples
presented in the previous section, for most of the helicity
configurations, the total run time required for obtaining full
analytic results for all the coefficients of the on-shell integrands
varies between a fraction of a second and about ten minutes, on a
single core.  For exceptionally complex cases (such as the one
involving functions of total degree up to 42), we need to perform the
reconstruction over up to three finite fields, for a total run time of
about a hundred minutes.

As a final observation, we point out that the purpose of the
calculations in sect.~\ref{sec:multi-loop-integrand} is to illustrate
a straightforward application of functional reconstruction algorithms
over finite fields to modern techniques for loop calculations.  In
particular, we have not attempted to find an integrand basis which
would yield simple results (such as the local integrands discussed in
ref.~\cite{Badger:2016ozq}).  We observe however that, using the
techniques we described in sect.~\ref{sec:mult-rati-funct}, one can
compute the total degree of the numerator and the denominator of the
results using a relatively small number of function evaluations.  This
can be used in order to estimate the complexity of the total result,
when looking for an integrand basis yielding simpler coefficients,
without the need of performing a full functional reconstruction.

\section{Conclusions and outlook}
\label{sec:conclusions-outlook}

We illustrated an algorithm for the reconstruction of multivariate
polynomials and rational functions from their evaluation over finite
fields, which has good scaling with the number of variables and the
complexity of the results.  We then discussed its application to
techniques for the calculation of scattering amplitudes in gauge
theories, such as the spinor-helicity formalism, tree-level recursion
relations, and multi-loop integrand reduction via generalized
unitarity.  The algorithm can compute complex analytic results,
side-stepping the issue of large intermediate expressions which would
arise in equivalent fully-analytic calculations.  The only input
required by the method is a procedure for the numerical evaluation
of the function to be reconstructed over finite fields.

As an application of these techniques, we presented for the first time
analytic results for the maximal cut of two five-point two-loop
topologies in Yang-Mills theory, for a complete set of independent
helicity configurations.  The complexity of the reconstructed results
highlights the suitability of the method to handle complex problems, in
particular high-multiplicity two-loop calculations, which are
currently of great relevance for high-energy phenomenology.

In this paper we discussed dense reconstruction algorithms.  Given
that some of our results contain several thousands of terms, we
believe this to be the best choice to handle the general case.
However, when the functions contain a relatively small number of terms
compared to the one expected from the total degree of the result,
dense algorithms can be outperformed by sparse algorithms.  A possible
approach is the usage of so-called racing algorithms, i.e.\ running a
dense and a sparse algorithm in parallel, sharing the same function
evaluations, and terminate the reconstruction when the fastest of the
two is successful.  If the time spent for the reconstruction is
dominated by the numerical evaluation of the function, as in our case,
this method can achieve optimal performance in virtually all cases.

One can observe that the simpler the result is, compared to the
intermediate expressions appearing in an equivalent analytic
calculation, the greater the advantage we can expect from using a
black-box reconstruction method rather than following a purely
analytic approach.  This makes the method ideal in the case where
large analytic cancellations yield simple final results.  Even for
somehow more complex results, such as many of the examples presented
in this paper, we found that our functional reconstruction
algorithm can easily outperform an equivalent analytic calculation by
several orders of magnitude.  We observed however that the analytic
results presented here might have been significantly simpler using a
better integrand basis, along the lines of what was proposed in
ref.~\cite{Badger:2016ozq}.  In particular, the capability of the
proposed reconstruction method of quickly computing the total degree
of the result can be exploited in order to estimate its complexity and
thus look for a choice of variables or an integrand basis which makes
the results simpler.  As a future application, the method can thus be
beneficial to the extension of the concepts outlined in
ref.~\cite{Badger:2016ozq} to a generic helicity configuration.

Although we focused on integrand reduction via generalized unitarity,
we stress that the number of potential applications of the proposed
approach is much larger.  Indeed, as we observed, any method which can
be numerically implemented via a sequence of elementary rational
operations is suited for the functional reconstruction algorithm we
illustrated.  On top of the ones presented here, other possible
applications are diagrammatic techniques and IBPs (indeed finite-field
applications to the latter have already been proposed in
ref.~\cite{vonManteuffel:2014ixa}).

The methods proposed in this paper are very general and have a very
broad spectrum of potential applications.  In particular, as shown by
our proof-of-concept implementation, they are suited for computing
complex analytic results in combination with state-of-the-art
multi-loop techniques.

\section*{Acknowledgements}
The author is indebted to Simon Badger for innumerable discussions and
suggestions, and in particular for pointing out the usefulness of
momentum-twistor variables for both analytic and numerical
calculations.  Thanks also go to Pierpaolo Mastrolia and Giovanni
Ossola for feedback on the draft and their ongoing collaboration on
integrand-reduction topics.  Several analytic results presented in
this paper have been numerically checked with a \textsc{Mathematica}
code developed by Simon Badger, Christian Brønnum-Hansen, and
Francesco Buciuni.  This work is supported by a Rutherford Grant
ST/M004104/1.

\begin{appendices}

\section{Basic finite-field algorithms}
\label{sec:basic-finite-field}

In this appendix we collect some basic algorithms relevant for the
finite-field applications presented in this paper.  These algorithms
are well known but not broadly used in high-energy physics, hence we
briefly describe them here for the convenience of the reader.

\subsection{Multiplicative inverse and rational reconstruction}
\label{sec:mult-inverse}

The calculation of a multiplicative inverse in $\Z_n$ and the rational
reconstruction algorithm both rely on the \emph{extended Euclidean
  algorithm}.  Given two integers $a,b\in\Z$, the extended Euclidean
algorithm yields their greatest common divisor $\gcd(a,b)$ and two
integers $s,t\in\Z$ such that
\begin{equation} \label{eq:exteucl}
  a\, s + b\, t = \gcd(a,b).
\end{equation}
It is useful to give a brief description of the algorithm, which
consists in generating the sequences of integers $\{r_i\}$, $\{s_i\}$,
$\{t_i\}$ and the integer quotients $\{q_i\}$ as follows
\begin{align}
  r_0={}&a\nn
  s_0={}&1\nn
  t_0={}&0\nn
  r_1={}&b\nn
  s_1={}&0\nn
  t_1={}&1\nn
  \cdots={}&\cdots\nn
  q_i ={} & \lfloor  r_{i-2}/r_{i-1}  \rfloor \nn
  r_i={}&r_{i-2}-q_i\, r_{i-1}\nn
  s_i={}&s_{i-2}-q_i\, s_{i-1}\nn
  t_i={}&t_{i-2}-q_i\, t_{i-1}.
\end{align}
Notice that, in practical implementations, only the last two entries
of each sequence need to be stored.  The algorithm terminates at $i=k$
when $r_k=1$, and identifies $t=t_{k-1}$, $s=s_{k-1}$, and
$\gcd(a,b)=r_{k-1}$.

Let us now turn to the calculation of a \emph{multiplicative inverse}
in $\Z_n$.  As we mentioned, a non-vanishing integer $a\in\Z_n$ admits
a multiplicative inverse if and only if $a$ and $n$ are co-prime,
i.e.\ $\gcd(a,n)=1$.  In this case, inserting $b=n$ in
Eq.~\eqref{eq:exteucl} and taking $\mmod n$ of both sides of the
equation gives
\begin{equation}
  a\, s \mod n = 1, \qquad \textrm{if $\gcd(a,n)=1$},
\end{equation}
which implies that we can identify
\begin{equation}
  a^{-1} \mod n = s, \qquad \textrm{if $\gcd(a,n)=1$}.  
\end{equation}
Because every non-vanishing element of a field must have a
multiplicative inverse, we can consider $\Z_n$ as a field if $n=p$ is
prime.  Notice that the calculation of the sequence $\{t_i\}$ is not
needed for the purposes of computing a multiplicative inverse.

A modified version of the extended Euclidean algorithm can also be
used for guessing a rational number $q$ from its image
$a\in\Z_n$~\cite{Wang:1981:PAU:800206.806398}.  Indeed, it is
straightforward to show that each iteration of the Euclidean algorithm
satisfies
\begin{equation}
  a\, s_i + b\, t_i = r_i,
\end{equation}
and therefore, by setting $b=n$ and taking $\mmod n$ of both sides,
\begin{equation}
  r_i = a\, s_i \mod n \quad \Leftrightarrow \quad  r_i/s_i \mod n = a.
\end{equation}
Hence $r_i/s_i$ is a possible guess for $q$ in each iteration of the
extended Euclidean algorithm.  However one can see that, when $n$ is
sufficiently large, most of these guesses will have very large
numerators and denominators, except for the case where
\begin{equation}
  r_i^2,s_i^2 \lesssim n.
\end{equation}
Hence the rational reconstruction algorithm can be implemented by
terminating the Euclidean algorithm at the iteration $i=k$, when
$r_k^2<n$.  The calculation of the sequence
$\{t_i\}$ is not needed for the rational reconstruction.

This algorithm typically succeeds in reconstructing a rational
number $q=a/b\in \Q$ from its image in $\Z_n$ when its numerator and
denominator satisfy
\begin{equation}
  a^2,b^2 \lesssim n.
\end{equation}
As explained in sect.~\ref{sec:implementation}, our requirement of
working with machine-size integers imposes an upper bound on our
choices of $n=p$ (in our implementation, $p<P_{\rm max}$ with
$P_{\rm max}=\O(10^6)$) and thus the former relation is generally not
guaranteed to hold.  However, as we mentioned in
sect.~\ref{sec:rati-reconstr}, this issue can be solved by performing
the functional reconstruction on several finite fields $\Z_{p_i}$ with
$p_i<P_{\rm max}$, and combining them by means of the Chinese remainder
theorem, as described below.

\subsection{Chinese remainder theorem}
\label{sec:chinese-remainder}
The Chinese remainder theorem allows us to reconstruct a number $a\in\Z_n$
with $n=n_1\cdots n_k$ and the $n_i$ pairwise co-prime, from its
images $a_i$ in the sets $\Z_{n_1},\ldots,Z_{n_k}$.  Hence, by
performing a calculation on several prime fields $\Z_{p_i}$, with
$p_i<P_{\rm max}$, and combining them with the Chinese remainder
theorem, one can reconstruct the image of the same result in
$\Z_{p_1\cdots p_k}$ and apply the rational reconstruction algorithm
described above to the latter.

More in detail, given $a\in Z_n$, a set of pairwise co-prime numbers
$n_1,\ldots n_k$ such that $n=n_1\cdots n_k$, and a set of congruences
\begin{equation}
  a_i = a \mmod{n_i},
\end{equation}
$a$ can be uniquely determined in $\Z_n$ as
\begin{equation}
  a = \sum_i m_i a_i \mmod{n},
\end{equation}
where
\begin{equation}
  m_i \equiv \left(\left(\frac{n}{n_i}\right)^{-1} \mmod{n_i}  \right)\, \frac{n}{n_i}.
\end{equation}

By using several primes $p_1,\ldots,p_k$, one can thus make their
product large enough for the reconstruction algorithm (described in
the previous section) to succeed in $\Z_{p_1\cdots p_k}$.  In
practice, as we explained, we perform the functional reconstruction
over one prime field $\Z_{p_i}$ at a time and recursively merge the
result on the bigger set $\Z_{p_1\cdots p_k}$.  The recursion
terminates when the result of the rational reconstruction is in
agreement with the evaluation of the function on different prime
fields.  Hence, all our calculations can be performed on $\Z_{p_i}$
with primes $p_i<P_{\rm max}$ using machine-size integers and only the
application of the Chinese remainder theorem requires multi-precision
arithmetic.

It is worth observing that, when applying the Chinese remainder
theorem to two congruences at a time, as in our case, one has
$n=n_1 n_2$ and the above formulas reduce to
\begin{equation}
  a = \big(m_1 a_1 + m_2 a_2\big) \mmod{n_1 n_2},
\end{equation}
with
\begin{equation}
  m_1 = (n_2^{-1} \mmod{n_1})\, n_2, \qquad  m_2 = (n_1^{-1} \mmod{n_2})\, n_1.
\end{equation}
The implementation can be further simplified by noting that
\begin{equation}
  m_2 = (1-m_1) \mmod{n_1 n_2}.
\end{equation}

\section{Six-dimensional momenta and spinors}
\label{sec:six-dimens-momenta}

We now briefly consider a six-dimensional generalization of some of
the techniques outlined in sect.~\ref{sec:four-dimens-momenta}.  For
the purposes of this paper, six-dimensional momenta and
spinors provide a higher-dimensional embedding of the loop
components in one- and two-loop calculations, which is used for the
applications of sect.~\ref{sec:multi-loop-integrand} in the context of
multi-loop generalized unitarity.

We use the following representation of a generic six-dimensional
momentum $p^\mu$, which is consistent with the light-cone
representation given in Eq.~\eqref{eq:4dmomenta} for four-dimensional
ones,
\begin{equation} \label{eq:6dmomenta}
  p \equiv (p^0+p^3,\ p^0-p^3,\ p^1+i\, p^2,\ p^1-i\, p^2,\ p^4+i\, p^5,\ p^4-i\, p^5).
\end{equation}
The six-dimensional spinor-helicity formalism has been extensively
developed in ref.~\cite{Cheung:2009dc}, to which we refer the reader
for details.  Here we only recall a few basic concepts useful
for our purposes.  In six dimensions, one can define four-components
spinors $|p^a\ra$ and $|p_{\dot a}]$ satisfying the Dirac equation
\begin{equation}
  p^\mu\, \sigma^{(6)}_\mu |p^a\ra = p^\mu\, \tilde \sigma^{(6)}_\mu |p_{\dot a}] = 0,
\end{equation}
where $\sigma_\mu^{(6)}$ and $\tilde \sigma_\mu^{(6)}$, for
$\mu=0,\ldots,5$, are now $4\times 4$ matrices which can be regarded
as higher-dimensional versions of the Pauli matrices (an explicit
representation can be found in ref.~\cite{Cheung:2009dc}).  The
indexes $a,\dot a = 0,1$ label the two independent solutions of the
Dirac equation, namely solutions with positive and negative helicity
for spinors and anti-spinors respectively.  A sub-set of the
six-dimensional spinor components can be identified with the ones of
$|p\ra$ and $|p]$ in the four-dimensional limit, hence making the
conversion from four to six dimensions trivial.  For our purposes, the
most important relation satisfied by six-dimensional spinors is
\begin{equation} \label{eq:sp6momenta}
  p_\mu = -\frac{1}{4}\, \la p^a | \sigma^\mu | p^b \ra \, \epsilon_{a b}, \qquad p_\mu = -\frac{1}{4}\, [ p_{\dot a} | \sigma^\mu | p_{\dot b} ] \, \epsilon^{\dot a \dot b},
\end{equation}
where $\epsilon^{a b}$ and $\epsilon_{\dot a \dot b}$ are
anti-symmetric Levi-Civita tensors in two dimensions with
$\epsilon^{01}=-\epsilon_{01}=1$, and a sum over repeated indexes is
understood.  The Levi-Civita tensors are also used for raising and
lowering the indexes $a$ and $\dot a$.

Unlike the four-dimensional case, where momenta are built from spinors
(which in turn are determined by their representation in terms of
momentum-twistor variables), in the six-dimensional one we are mostly
interested in the opposite, i.e.\ building spinors $|p^a\ra$ and
$|p_{\dot a}]$ from the entries in Eq.~\eqref{eq:6dmomenta} of a
six-dimensional momentum, such that Eq.\eqref{eq:sp6momenta} is
satisfied.  This is because, in the context of generalized unitarity,
we start from six-dimensional momenta corresponding to on-shell loop
propagators, and from these we must build their corresponding spinors.
While the solution of the problem is not unique, it is not hard to
work out a suitable representation, although one has to separately
consider special cases where one or more of the entries are vanishing.
It should be pointed out that, once the sum over the internal
helicities of a unitarity cut has been performed, spinors
corresponding to loop momenta always combine as in the
right-hand-sides of Eq.~\eqref{eq:sp6momenta}, hence these relations
(and the Dirac equation) are the only relevant ones for this purpose.

The last six-dimensional ingredient we need are polarization vectors,
for cut loop propagators involving gluons.  For six-dimensional gluons
we have four independent polarizations, which can be identified by the
labels $(a\, \dot a)$ of their $SU(2)\times SU(2)$ representation
\begin{equation}
  (a\, \dot a) \in \{(00), (11), (01), (10)\} \equiv \{(++), (--), (+-), (-+)\}.
\end{equation}
While $(++)$ and $(--)$ respectively correspond to the positive and
negative helicity in the four-dimensional limit, the last two are
instead only present in six dimensions.  Given an auxiliary vector
$\eta^\mu$, polarization vectors can be written
as~\cite{Cheung:2009dc}
\begin{equation}
  \epsilon^\mu_{a \dot a}(p,\eta) = \frac{1}{\sqrt{2}\, (p\cdot \eta)}\, \la p_a | \sigma^\mu | \eta_{b} \ra\, \la \eta_c | p _{\dot a} ]\, \epsilon^{b c}.
\end{equation}
As in the four-dimensional case, in our representation we divide them
by an additional factor $\sqrt{2}$, such that their light-cone
components become rational functions of the spinor variables.  In
ref.~\cite{Cheung:2009dc} it is shown that these vectors obey all the
properties required by polarizations of gauge bosons, and the
following relation
\begin{equation} \label{eq:6depssum}
  \epsilon^\mu_{a \dot a}(p,\eta)\, \epsilon^\nu{}^{a \dot a}(p,\eta) = g^{\mu \nu} - \frac{1}{(p\cdot \eta)} \left( p^\mu \eta^\nu + p^\nu \eta^\mu \right).
\end{equation}
The latter is important in the context of generalized unitarity.
Indeed, after the sum over the internal helicities, polarization
vectors corresponding to cut loop propagators always appear in the
combination on the l.h.s.\ of the equation.

Similarly to the four-dimensional case, the ingredients reviewed here
allow us to work with six-dimensional spinors by performing rational
operations on their components, hence they are suited for numerical
evaluations over finite fields $\Z_p$ and thus for the application of
the functional reconstruction algorithms described in
sect.~\ref{sec:funct-reconstr}.

\section{Two-loop unitarity cuts from Berends-Giele currents}
\label{sec:two-loop-unitarity}

In this appendix we briefly illustrate how to evaluate
two-loop unitarity cuts efficiently using off-shell Berends-Giele currents.  This
is a straightforward generalization of the algorithm used by the
public numerical code \textsc{NJet}~\cite{Badger:2012pg} at one-loop.
The algorithm is suited for numerical implementations, including
evaluations over finite fields $\Z_p$.

\begin{figure}[t]
  \centering
  \includegraphics[width=0.6\textwidth]{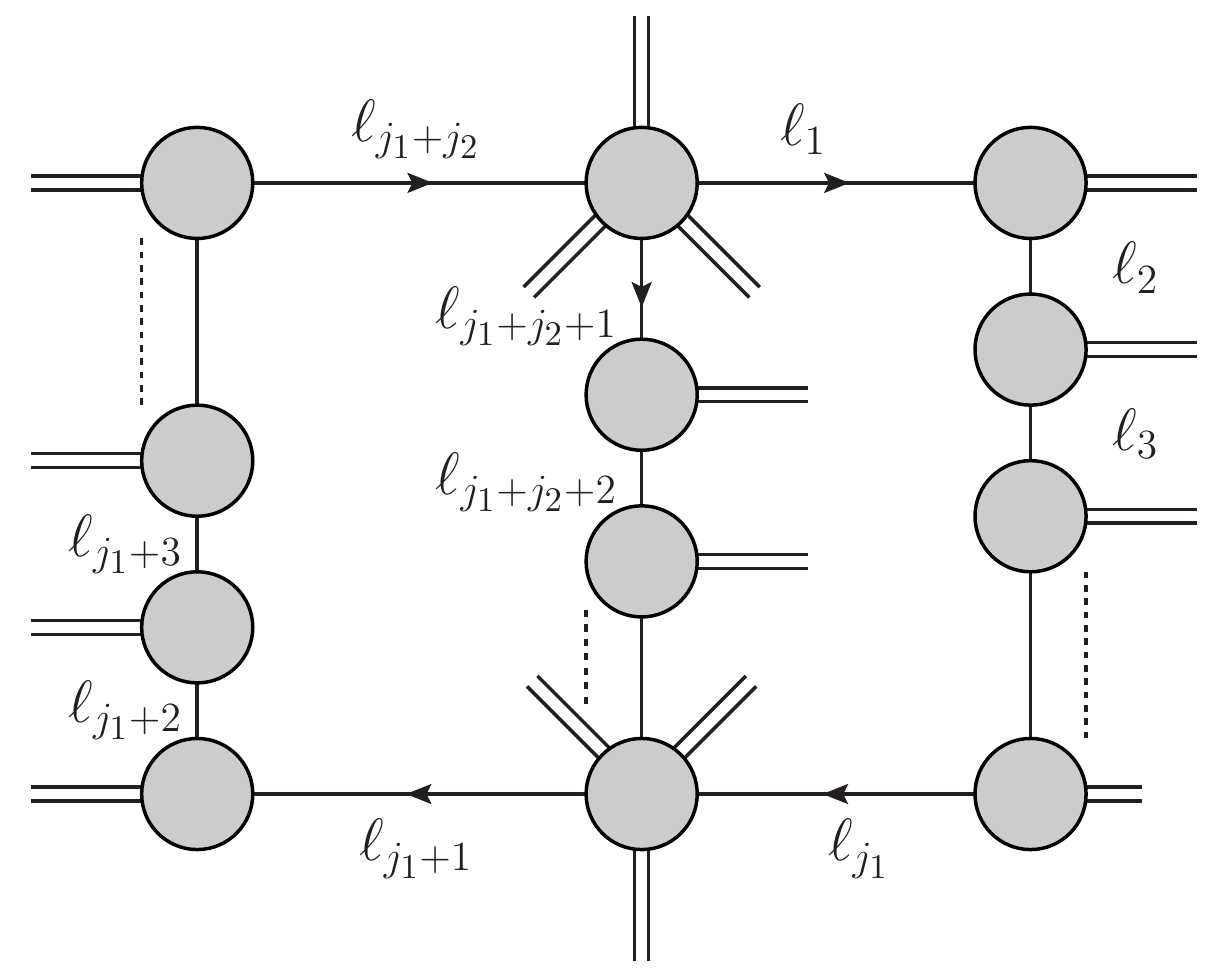}
  \caption{Schematic depiction of a unitarity cut.  Grey blobs
    represent tree-level amplitudes and they are joined by the lines
    corresponding to the on-shell momenta of the cut loop propagators
    $\ell_i$.  The loop momenta are defined as $k_1=\ell_1$, and
    $k_2=-\ell_{j_1+j_2}$.  Double lines represent an arbitrary number
    of external legs.}
  \label{fig:loopcurrents}
\end{figure}

For simplicity, we consider a multiple cut of the form depicted in
fig.~\ref{fig:loopcurrents}, for a theory with only gluons, but it
should be clear that everything can be easily extended to more general
cases.
In particular, fig.~\ref{fig:loopcurrents} represents a unitarity cut
where the momenta of the loop propagators which are put on-shell are
denoted by $\ell_i$.  We split the loop propagators in three
categories: $\{\ell_{1},\ldots,\ell_{j_1}\}$ are propagators depending
on the loop momentum $k_1$ only,
$\{\ell_{j_1+1},\ldots,\ell_{j_1+j_2}\}$ depend on $k_2$ only, and
$\{\ell_{j_1+j_1+1},\ldots,\ell_{j_1+j_2+j_{12}}\}$ depend on both
$k_1$ and $k_2$.  Double lines represent an arbitrary set of external
legs.  The cut is defined as the product of the tree-level amplitudes
(represented as grey blobs in the picture) defined by the on-shell
propagators, summed over the helicity states of the internal legs.

We first focus on the tree-level amplitude involving $\ell_1$ and
$\ell_2$.  When this amplitude is computed via Berends-Giele
recursion, the last step of such recursion is the contraction of the
on-shell current involving $\ell_1$ and the appropriate sequence of
external legs with the polarization vector corresponding to the
on-shell propagator $\ell_2$.  This is schematically represented by
the following equation
\begin{equation} \label{eq:loopfirsttree}
  \includegraphics[scale=0.5, trim=0 8 0 0]{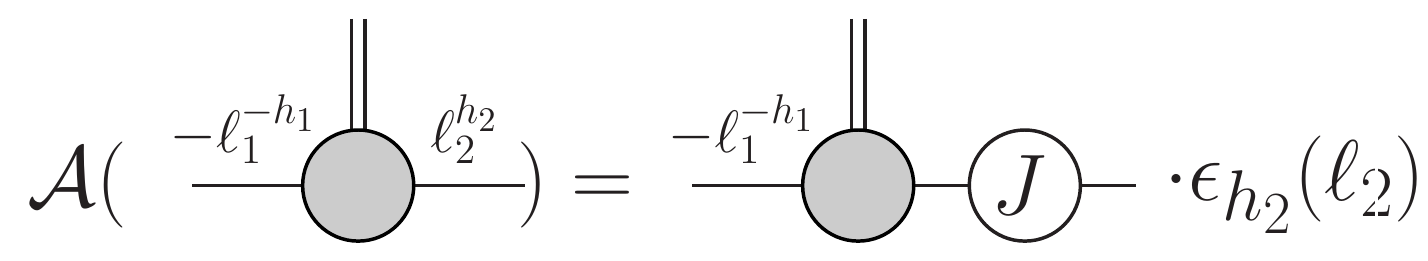}.
\end{equation}
In the calculation of the current appearing on the r.h.s.\ of the
previous equation, the lower-point currents depending only on the
external legs only need to be computed once per phase-space point,
since they are independent of the cut.  We now recall that, in the
product of amplitudes defined by this unitarity cut, the polarization
vector $\epsilon_{h_2}(\ell_2)$ always appears in the combination
$\epsilon_{h_2}^\mu(\ell_2) \epsilon_{-h_2}^\nu(-\ell_2)$.  After the
sum over internal helicities we have
\begin{equation}
  \sum_{h}{\textrm{sign}(h_2)}\, \epsilon_{h}^\mu (\ell_2)\, \epsilon_{-h}^\nu(-\ell_2) = g^{\mu\nu} - \frac{1}{(\ell_2\cdot \eta)} \left( \ell_2^\mu \eta^\nu + \ell_2^\nu \eta^\mu \right),
\end{equation}
as apparent e.g.\ from Eq.~\eqref{eq:6depssum}.  In the previous
equation, $\eta$ is the reference vector used to define the internal
polarizations, and the second term on the r.h.s.\ may be dropped
because of gauge invariance.  This implies that, instead of computing
a product of amplitudes and summing over the internal helicities, one
can equivalently simply compute the Berends-Giele current on the
r.h.s.\ of \eqref{eq:loopfirsttree}, without contracting it with the
polarization vector $\epsilon_{h_2}(\ell_2)$, and use this current as
an input on-shell current for the calculation of the next amplitude
(in this case, the one involving $\ell_2$ and $\ell_3$), as if the
current itself was the polarization vector associated to $\ell_2$.

The argument can obviously be iterated over all propagators
$\ell_i$ with $i\leq j_1$, i.e.\ those which depend on the first loop
momentum $k_1$ only, so that for each helicity $h$ associated with
$\ell_1$ we build a current $J_{k_1,h}$ as
\begin{equation} \label{eq:jl1currents}
  \includegraphics[scale=0.5, trim=0 8 0 0]{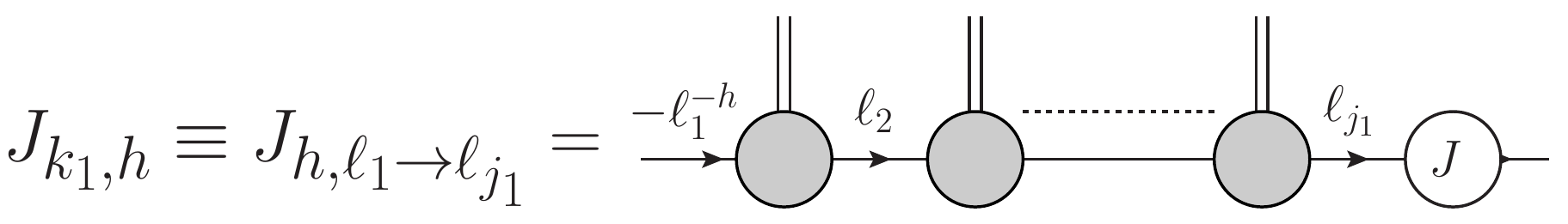},
\end{equation}
where each sum over the internal helicities of cut propagators is
replaced by a contraction with $g^{\mu \nu}$.  As already observed,
lower-point currents depending on external legs only need to be
computed once per phase-space point and can be reused on different cuts.
Similarly, for each helicity $h$ associated with $\ell_{j_1+1}$ in
fig.~\ref{fig:loopcurrents} we build a current for the propagators
depending on $k_2$ only, as
\begin{equation} \label{eq:jl2currents}
  \includegraphics[scale=0.5, trim=0 8 0 0]{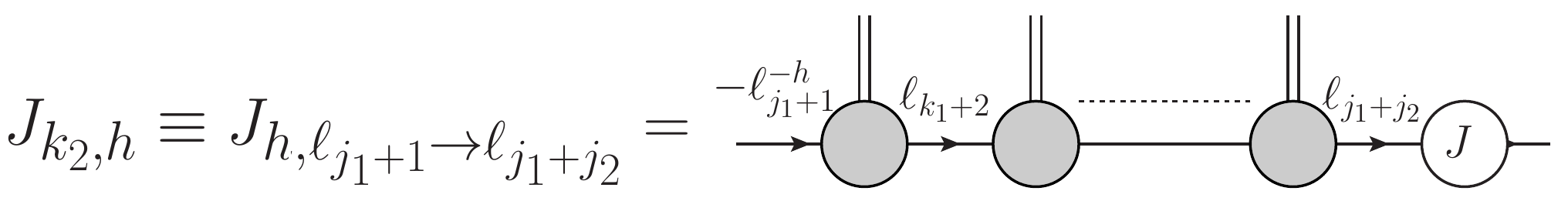}.
\end{equation}
At this point, we start building currents depending on both $k_1$ and
$k_2$, using $J_{k_1,h}$ and $J_{k_2,h}$ as the input on-shell
currents of the recursion.  In particular, we define the currents
$J_{\textrm{up},k_1,k_2,h_1,h_2}$ and
$J_{\textrm{down},k_1,k_2,h_1,h_2}$ corresponding to the upper and
lower intersection of the two loops in fig.~\ref{fig:loopcurrents},
namely
\begin{equation} \label{eq:jupjdowncurrents}
  \includegraphics[scale=0.5, trim=0 8 0 0]{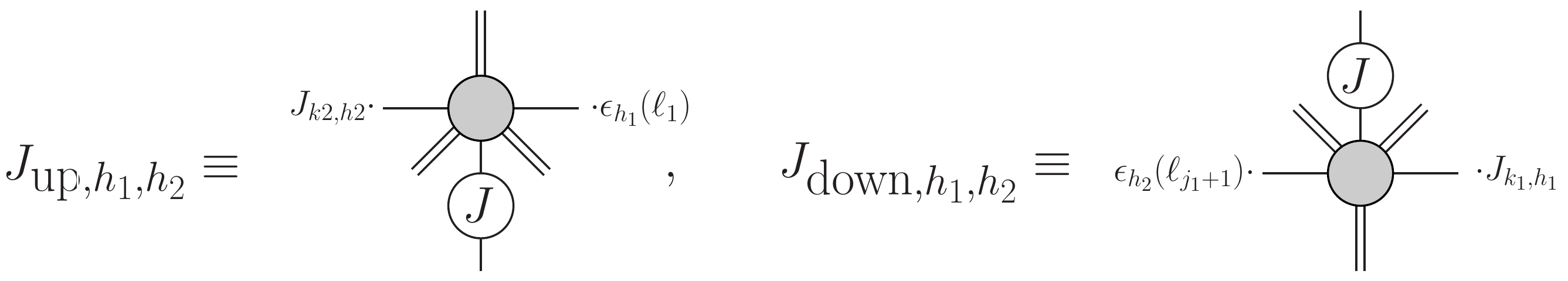}.
\end{equation}
Using $J_{\textrm{up},k_1,k_2,h_1,h_2}$ as input, we move down along
the third sequence of propagators $\ell_{j_1+j_2+1}$,
$\ell_{j_1+j_2+2}$,\ldots\ depending on both $k_1$ and $k_2$, by means
of the same algorithm which defined the currents $J_{k_1,h}$ and
$J_{k_2,h}$,
\begin{equation} \label{eq:jl12currents}
  \includegraphics[scale=0.5, trim=0 8 0 0]{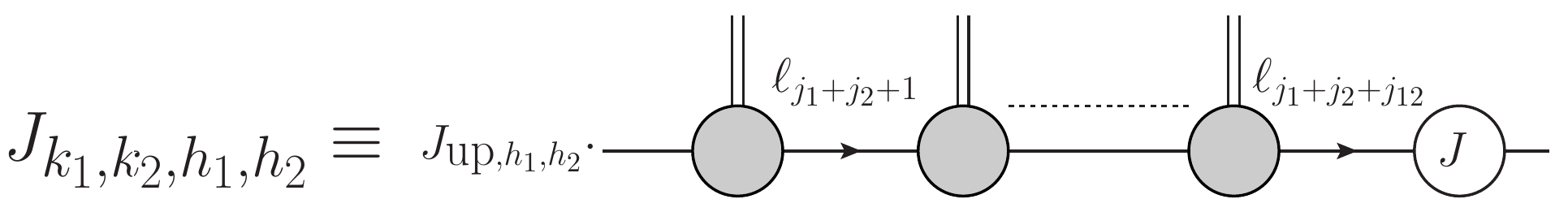}.
\end{equation}
The value of the cut is thus
\begin{equation}
  \sum_{h_1 h_2}\, \textrm{sign}(h_1)\, \textrm{sign}(h_2)\, J_{k_1,k_2,h_1,h_2}\cdot J_{\textrm{down},k_1,k_2,h_1,h_2}.
\end{equation}

\end{appendices}

\bibliographystyle{JHEP}
\bibliography{biblio}

\end{document}